\documentclass[preprint,aps,12pt,nofootinbib,superscriptaddress]{revtex4}
\pdfoutput=1
\usepackage[pdftex]{color,graphicx}
\usepackage{mathrsfs}
\usepackage{amssymb}
\usepackage{amsmath}
\usepackage{mathtools}
\usepackage{epsfig}
\usepackage{graphics}
\usepackage{cancel}
\usepackage{bbm}
\usepackage{slashed}

\newcommand{\be}{\begin{equation}}
\newcommand{\ee}{\end{equation}}
\newcommand{\bea}{\begin{eqnarray}}
\newcommand{\eea}{\end{eqnarray}}

\newcommand{\ba}{\begin{array}}
\newcommand{\ea}{\end{array}}

\begin{document}

\title{On neutrinoless double beta decay in the minimal left-right symmetric model}

\author{Wei-Chih Huang}
\email[]{whuang@sissa.it}
\affiliation{SISSA, via Bonomea 265, 34136 Trieste, Italy.}
\affiliation{INFN, sezione di Trieste, 34136 Trieste, Italy.}
\affiliation{Department of Physics and Astronomy, University College London,\\ London WC1E 6BT, United Kingdom}
\author{J. Lopez-Pavon}
\email[]{jlpavon@sissa.it}
\affiliation{SISSA, via Bonomea 265, 34136 Trieste, Italy.}
\affiliation{INFN, sezione di Trieste, 34136 Trieste, Italy.}

\begin{abstract}
We analyze the general phenomenology of neutrinoless double beta decay in the minimal left-right symmetric
model. We study under which conditions a New Physics dominated neutrinoless double beta
decay signal can be expected in the future experiments. We show that the correlation
among the different contributions to the process, which arises from the neutrino mass generation mechanism,
can play a crucial role. We have found that, if no fine tuned cancellation is involved in the light active neutrino contribution, 
a New Physics signal can be expected mainly from the $W_R-W_R$ channel. An interesting exception is the 
$W_L-W_R$ channel which can give a dominant contribution to the process if the right-handed
neutrino spectrum is hierarchical with $M_1\lesssim$ MeV and $M_2,M_3\gtrsim$ GeV.
We also discuss if a New Physics signal in neutrinoless double beta decay experiments is 
compatible with the existence of a successful Dark Matter candidate in the left-right symmetric 
models. It turns out that, although it is not a generic feature of the theory, it is 
still possible to accommodate such a signal with a KeV sterile neutrino as Dark matter.
\end{abstract}

\preprint{SISSA 41/2013/FISI}

\maketitle

\section{Introduction}

The recent LHC results~\cite{Aad:2012tfa,Chatrchyan:2012ufa} seem to indicate that the Higgs mechanism, with
the Higgs mass around $125$ GeV, is the responsible for the mass generation of the Standard Model (SM) particles.
However, the origin of light neutrino masses, for the existence of which we have compelling evidences from
neutrino oscillation experiments, still remains unknown. It is true that the light neutrino masses
could also be generated through the Higgs mechanism in a minimally extended SM which includes
sterile (right-handed) neutrino fields as $SU(2)_L$ singlets and in which the total lepton number
is conserved. However, their smallness in comparison with the charged lepton and quark masses
calls for a different explanation. In this context, extensions of the SM required to explain
the origin of neutrino masses, and compatible with the latest LHC data, arise as
quite suggestive models of New Physics (NP). Among those we find the celebrated seesaw
models~\cite{Minkowski:1977sc,GellMann:1980vs,Yanagida:1979as,Mohapatra:1979ia},
which can give us, in addition, the key to explain the matter-antimatter asymmetry of the universe
through Leptogenesis~\cite{Fukugita:1986hr}.

Most of those models predict that neutrinos are Majorana particles,
something which can be tested in lepton number violating processes such as the neutrinoless
double beta ($0\nu\beta\beta$) decay. The $0\nu\beta\beta$ decay experiments are the most promising
ones in this context but they suffer a serious drawback: the NP contribution to the process is usually short 
range and thus typically very suppressed compared to that of the light neutrinos. Thanks to  
the future $0\nu\beta\beta$ experiments~\cite{CUORE,Auger:2012ar,GERDA,KamLANDZen:2012aa,Gaitskell:2003zr,Diaz:2009zzb,Barabash:2004pp},
in combination with the complementary information coming from neutrino oscillation experiments
and cosmology, we might be able to discover the Majorana nature of neutrinos, but not easily 
which is the mechanism responsible for the neutrino mass generation~\cite{Faessler:2011qw,Meroni:2012qf}. In this context,
the correlations between the standard light neutrino and NP contribution to the $0\nu\beta\beta$ decay are
crucial, as shown in the case of the type-I~\cite{Minkowski:1977sc,GellMann:1980vs,Yanagida:1979as,Mohapatra:1979ia},
type-II~\cite{Magg:1980ut,Schechter:1980gr,Wetterich:1981bx,Lazarides:1980nt,Mohapatra:1980yp} and
type-III~\cite{Foot:1988aq} seesaw models in Refs.~\cite{LopezPavon:2012zg,Blennow:2010th}. The generation
of light neutrino masses in a particular model usually induces important correlations between the
different contributions to the $0\nu\beta\beta$ decay, which should always be considered in a model 
dependent analysis, helping to understand which type of NP can be feasibly tested in the experiments.

In this work we will
focus on the $0\nu\beta\beta$ decay phenomenology of the minimal left-right symmetric model 
(MLRSM)~\cite{Pati:1974yy,Mohapatra:1974gc,Senjanovic:1975rk,Mohapatra:1979ia, Mohapatra:1980yp}.
The left-right symmetric models have been widely studied in the literature since, among other features, they provide a
natural explanation for the smallness of the neutrino masses (some recent analysis in the context of the $0\nu\beta\beta$ decay can be
found in Refs.~\cite{Nemevsek:2011aa,Tello_thesis:2012,Chakrabortty:2012mh,Nemevsek:2012iq,Parida:2012sq,Barry:2013xxa,Dev:2013vxa,Dev:2013oxa}).
In our analysis we will assume that no accidental cancellation occurs in the light neutrino mediated 
$W_L-W_L$ channel, which involves the exchange of two $W_L$. We will distinguish three regions of the parameter space depending on the 
mass of the right-handed (RH) neutrinos. First, we will show that if the right-handed (RH) neutrinos
are heavier than the $0\nu\beta\beta$ decay scale ($\sim100$ MeV), the $0\nu\beta\beta$ decay rate is dominated by light
neutrino exchange channels with the exception of the channel in which two $W_R$ are exchanged ($W_R-W_R$ channel)
mediated by heavy neutrinos\footnote{If other contributions coming from different models are not involved.}.
One of the light neutrino mediated channels involves the exchange of one $W_L$ and one $W_R$ ($W_L-W_R$ channel); however, it turns out that a NP dominant contribution can come
mainly from the $W_R-W_R$ channel.
Secondly, we will study the region of the parameter space where the RH neutrinos are lighter than the 
$0\nu\beta\beta$ scale. We have found that in this case the $W_L-W_R$ contribution cancels out while a 
NP signal can still be expected from the $W_R-W_R$ channel. In this region, the RH 
neutrinos can give a relevant contribution through the $W_L-W_L$ channel, as opposed to the type-I seesaw 
case where the total $W_L-W_L$ contribution is very suppressed. Finally,
we will investigate a mixed scenario with RH neutrinos in both regions below and above the $0\nu\beta\beta$
decay scale. We have found that this is the only scenario in which the $W_L-W_R$ channel turns out to be relevant and can be 
responsible of a future signal (if no cancellation in the $W_L-W_L$ channel is invoked). In all the cases we 
will show for which part of the parameter space a NP signal in future $0\nu\beta\beta$ decay experiments 
can be expected. Moreover, we will also analyze if such a signal can be compatible with the existence of 
a successful Dark Matter (DM) candidate in the left-right symmetric model, study the complementary
bounds coming from charged lepton flavour violation (LFV) experiments and the impact of the 1-loop corrections 
to the light neutrino masses.

This work is organized as follows. In Section 2 we briefly describe the MLRSM,
focusing on the relations among the parameters of the model induced by the neutrino mass
generation. In Section 3 we analyze the neutrinoless double beta decay phenomenology in the MLRSM,
studying in particular for which part of the parameter space a $0\nu\beta\beta$ decay
signal coming mainly from NP contributions can be possible. Section 4 is devoted to the analysis 
of complementary constraints coming mainly from charged LFV experiments and the 
stability of the light neutrino masses under 1-loop corrections. In Section 5 we study if a successful 
DM candidate and a NP signal in the future $0\nu\beta\beta$ decay experiments can be compatible in 
the MLRSM. Finally, we conclude in Section 6.

\section{Minimal left-right symmetric model and neutrino masses}

The Lagrangian of the MLRSM respects an enlarged gauge symmetry $SU\left(3\right)_c\otimes SU\left(2\right)_L\otimes SU\left(2\right)_R\otimes
U\left(1\right)_{B-L}$ plus a discrete left-right symmetry which leads to equal $SU\left(2\right)_L$ and $SU\left(2\right)_R$
gauge couplings ($g_L=g_R=g$). We are not going into the details of the model since it has been widely studied in
the literature (for a recent complete analysis regarding the associated lepton number violating effects, see for instance
Ref.~\cite{Tello_thesis:2012,Barry:2013xxa}), but only recall the most relevant features for
our analysis. The scalar sector is also augmented by the addition of two scalar triplets ($\Delta_L$ and $\Delta_R$)
and a bi-doublet scalar under $SU(2)_L \otimes SU(2)_R$, which spontaneously break the electroweak symmetry when they develop vacuum
expectation values (vevs).

In this section we will derive the relations which will be used in the phenomenological analysis of the
$0\nu\beta\beta$ decay. Since they come from the neutrino mass generation, let us recall how
the complete neutrino mass matrix looks like after the electroweak symmetry breaking:

\be
 M_\nu = \begin{pmatrix} M_L &  m_D^T \\
 m_D & M_R  \end{pmatrix} = U \,\text{Diag}\left(m,M\right)U^T,
\label{Mnu}
\ee
where $m_i$ are the light neutrino masses and $M_i$ the heavy ones. Notice
that in this model the Majorana mass term for the heavy neutrinos is generated dynamically when $\Delta_R$
takes a vev ($M_R=Y_{\Delta_R} v_R$), while the Majorana mass term $M_L$ for the left-handed (LH) neutrinos is
generated analogously through the $\Delta_L$ vev ($M_L = Y_{\Delta_L} v_L$). The neutrino mass matrix
is diagonalized as shown above by a $6\times 6$ unitary matrix $U$, through the following rotation between the neutrino flavor and mass eigenstates
denoted by $\alpha,\beta = e, \mu, \tau$ and $i,k=1,2,3$, respectively,

\be
\begin{pmatrix}
\nu_{\alpha L} \\ N^c_{\beta R} \end{pmatrix} = U \begin{pmatrix}
\nu_{i} \\ N_{k} \end{pmatrix}=\begin{pmatrix}
\tilde{U} & B \\ A & V \end{pmatrix}
\begin{pmatrix}
\nu_{i} \\ N_{k}
\end{pmatrix},
\ee
The diagonalization of the complete neutrino mass matrix presented in
Eq.~(\ref{Mnu}) provides the following useful relations

\bea
\label{constraintsa}
\tilde{U}m\tilde{U}^T + BM B^T&=& M_L,\\
\label{constraintsb}
\tilde{U}mA^T + BMV^T &=& m_D^T,\\
\label{constraintsc}
AmA^T + VMV^T &=& M_R,
\eea

On the other hand, taking into account that the active LH block of $U$, $\tilde{U}$, is unitary to a
very good approximation (at least up to the percent level~\cite{Antusch:2006vwa}), the complete neutrino mixing matrix can be expanded as,

\be
U =\begin{pmatrix} 1-\theta\theta^\dagger/2 & \theta \\
 -\theta^\dagger & 1-\theta^\dagger \theta/2 \end{pmatrix} \begin{pmatrix} U_{pmns} & 0\\
 0 & V \end{pmatrix} +\mathcal{O}\left( \theta^3 \right) = \begin{pmatrix} U_{pmns} & \theta V\\
 -\theta^\dagger U_{pmns} & V \end{pmatrix} +\mathcal{O}\left( \theta^2 \right),
\label{U}
\ee
where $\theta$ is a $3\times 3$ matrix which characterizes the small mixing
between the active LH and the heavy RH neutrinos, $U_{pmns}$ is the PMNS matrix and $V$ is a $3\times3$ unitary matrix. From
Eqs.~(\ref{constraintsa}-\ref{constraintsc}), we have

\bea
U_{pmns}\,m\,U_{pmns}^T &=& M_L-\theta M_R \theta^T,\label{constraints1}\\
\theta M_R - M_L\theta^* &=& m_D^T,\label{constraints2}\\
VMV^T &=& M_R\left( 1+\mathcal{O}\left(\frac{M_L}{M_R}\theta^2\right)\right).
\label{constraints3}
\eea
The discrete (charge conjugation) LR symmetry gives us the following relation between the Yukawa couplings of the triplets:
$Y_{\Delta_R}=Y_{\Delta_L}\equiv Y_\Delta$.\footnote{Another option is to consider instead a discrete parity
symmetry leading to a similar relation: $Y_{\Delta_R}=Y_{\Delta_L}^*\equiv Y_\Delta$.} This means that

\be
(M_L)_{\alpha\beta}/(M_R)_{\alpha\beta}=v_L/v_R < 10^{-3},
\ee
where we have employed the present bounds on $v_L$ and $v_R$, namely $v_L\lesssim 7$
GeV~\cite{Melfo:2011nx} and $v_R \gtrsim 10$ TeV
($M_{W_R}\approx gv_R/\sqrt{2}\gtrsim$ TeV~\cite{Beringer:1900zz,Abazov:2011xs,Aad:2012ej}). Therefore,
the $\mathcal{O}\left( M_L\theta^*  \right)$ and $\mathcal{O}\left(\frac{M_L}{M_R}\theta^2\right)$
can be safely neglected in Eq.~(\ref{constraints2}) and Eq.~(\ref{constraints3}) respectively, and

\be
\theta \simeq m_D^T M_R^{-1}.
\label{theta}
\ee
Of course, $\theta$ plays a fundamental role at the phenomenological level since it basically
describes the mixing between the active LH neutrinos and the RH ones. It would be
very interesting thus to find a useful parametrization of $\theta$ as a function of the light neutrino
parameters, light neutrino masses and the angles/phases of the PMNS matrix, and the rest of the
independent parameters of the model associated with the RH neutrino sector. In principle, an
analogous parametrization to the Casas-Ibarra one~\cite{Casas:2001sr} would be a good 
candidate~\cite{Akhmedov:2008tb}. However, the presence of $M_L$ in Eq.~(\ref{constraints1}) and the fact 
that the matrix $V$ is in this case physical, contrary to the type-I seesaw model, makes that 
parameterization less transparent and more involved than expected. On the other hand, the discrete 
(charge conjugation) LR symmetry leads to the following constraint

\be
m_D=m_D^T,
\ee
and thus, Eq.~(\ref{theta}) becomes

\be
\theta M_R = M_R \theta^T=m_D.
\ee
Plugging this relation and Eq.~(\ref{constraints3}) into Eq.~(\ref{constraints1}), we obtain,

\be
U_{pmns}\,m\,U_{pmns}^T = M_L-\theta^2 VMV^T,
\ee
and finally with $Y_{\Delta_R}=Y_{\Delta_L}\equiv Y_\Delta$ and hence $M_L=\frac{v_L}{v_R}M_R$,
we have,

\be
\theta= \left[\frac{v_L}{v_R}I-U_{pmns}\,m\,U_{pmns}^TV^*M^{-1}V^\dagger \right]^{1/2}.
\label{theta2}
\ee
Therefore, $\theta$ is completely determined as a function of the light and heavy neutrino masses, $m$ and $M$, the
PMNS matrix, $U_{pmns}$, $v_L/v_R$, and the unitary matrix $V$~\cite{Nemevsek:2012iq}. Notice that if
this expression is used to obtain $\theta$ with the PMNS mixing angles and the solar and atmospheric
mass-squared differences as input parameters, we ensure that the model is consistent with the light neutrino mass
and mixing pattern measured in neutrino oscillation experiments.

\section{Neutrinoless double beta decay}

In our study of the $0\nu\beta\beta$ decay in the MLRSM, we will pay special attention to the correlation among all
the contributions to the process and, in particular, the connection with the light neutrino masses. We shall
see that the correlation between the different contributions and the experimental bounds on the parameters will
allow us to safely neglect some of the NP contributions.

As we have already mentioned, we will not analyze the scenario in which a cancellation
occurs within the standard light neutrino contribution, which would naively
leave the NP channels as the leading contributions~\cite{Ibarra:2011xn,Mitra:2011qr}. Of course, this cancellation can be
due to the presence of an extra symmetry added to the model, such as the
lepton number which is approximately conserved in the so called inverse or direct seesaw
models~\cite{Mohapatra:1986bd,Branco:1988ex,Malinsky:2005bi,Shaposhnikov:2006nn,Gavela:2009cd}. The problem in this
scenario is that, in order for the NP contributions to be measurable, a significant violation of lepton number should be introduced
through the NP sector which may not have an impact on the light neutrino masses at tree level but arises
naturally at one-loop level, as shown in Ref.~\cite{LopezPavon:2012zg} in the context of the seesaw models.
This makes it very difficult to have a significant contribution from NP channels since the 1-loop
correction to the light neutrino masses tends to dominate in the $0\nu\beta\beta$ decay rate.

We will distinguish three different regions according to the associated $0\nu\beta\beta$ decay phenomenology: (i)  when the
RH neutrinos are much heavier than the $0\nu\beta\beta$
decay scale ($\langle p\rangle \approx 100$ MeV), which means heavier than approximately $1$ GeV; (ii) when the RH
neutrinos are much lighter than the $0\nu\beta\beta$ decay scale (below $1$ MeV); (iii) when the RH neutrinos are in both regions, (i) and (ii).

\begin{figure}
\begin{center}
 \includegraphics[width=1\textwidth,angle=0]{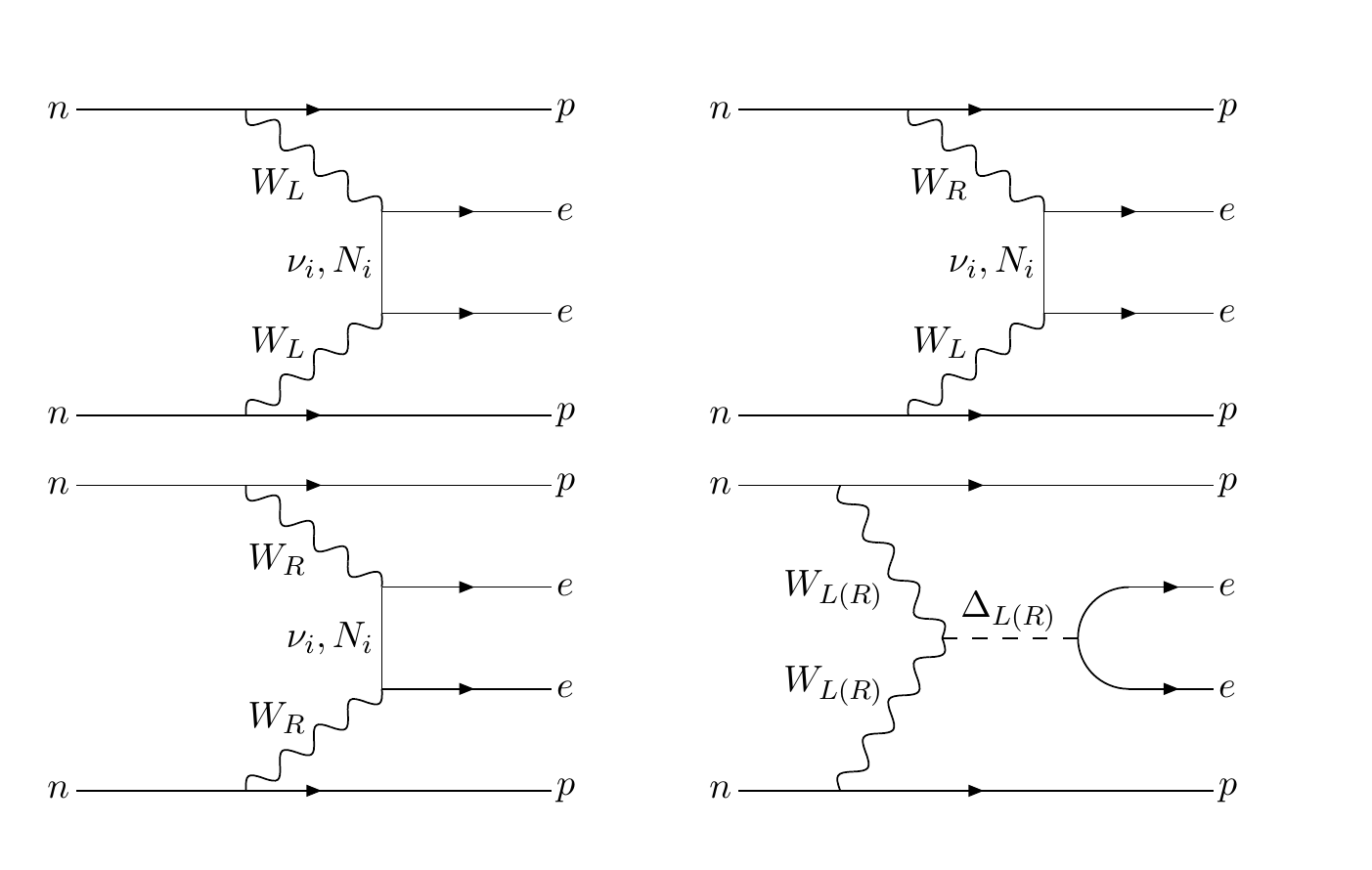}
\end{center}
\caption{Feynman diagrams contributing to the $0\nu\beta\beta$ transition rate in the MLRSM.}
\label{feyn}
\end{figure}

In the analysis below we have reasonably estimated the NMEs corresponding to 
some of the channels under study. This is accurate enough for our purposes but, although the associated NMEs 
errors are still large, in order to be more precise, full calculation of all the NMEs should 
be considered. 

\subsection{Heavy regime}
\label{heavyregime}
The various contributions to the $0\nu\beta\beta$ transition rate in this model are described by the
Feynman diagrams shown in Fig.~\ref{feyn}. We will start describing them one by one in order to show that the
contributions of the diagrams in which a heavy fermion (or scalar) is exchanged are subdominant with respect to
those of the light neutrino exchange, the only exception being the $W_R-W_R$ channel.

\begin{itemize}

\item
$\mathbf{W_L-W_L}$ \textbf{channel}. The amplitude corresponding to the top left diagram of Fig.~\ref{feyn} is given by:

\be
A_{LL} \propto \left(1+\mathcal{O}\left(\xi\right)\right)\left[\sum_{i=1}^3 m_i \tilde{U}_{ei}^2 \mathcal{M}^{0\nu\beta\beta}(0)+
\sum_{i=1}^3M_i B_{ei}^2 \mathcal{M}^{0\nu\beta\beta}(M_i)\right],\nonumber
\ee
where $\xi$ is the $W_L-W_R$ mixing angle. The present experimental bound is given by 
$\xi\lesssim10^{-2}$~\cite{Beringer:1900zz}, however, in the minimal left-right symmetric model 
there is a stronger theoretical upper bound given by $M_{W_L}^2/M_{W_R}^2<10^{-3}$~\cite{Aad:2012ej}.
\footnote{Notice that $\xi$ can only saturate this bound if and only if the two vev's of the 
Higgs doublets are of the same order.} In the case of the $W_L-W_L$ contribution, $\xi$ can be
safely neglected. $\mathcal{M}^{0\nu\beta\beta}$ are the associated nuclear matrix elements (NMEs) following the notation
of Ref.~\cite{Blennow:2010th}, where the NMEs were computed as a function of
the mass of the neutrino mediating the process for different nuclei. Notice that in this notation the NMEs
include the dependence on the propagator. The NMEs corresponding to the light neutrino exchange are independent
of the neutrino masses, and then with Eq.~(\ref{constraintsa}), the above amplitude can be rewritten as
\be
A_{LL} \propto (M_L)_{ee} \mathcal{M}^{0\nu\beta\beta}(0)+
\sum_{i=1}^3M_i B_{ei}^2 \left(\mathcal{M}^{0\nu\beta\beta}(M_i)-\mathcal{M}^{0\nu\beta\beta}(0)
\right).
\label{ALL}
\ee
Taking into account that $\mathcal{M}^{0\nu\beta\beta}(M_i)/\mathcal{M}^{0\nu\beta\beta}(0)\ll1$~\cite{nme},
the contribution due to the heavy neutrino exchange can be safely neglected. Using again
Eq.~(\ref{constraintsa}), one obtains,
\be
A_{LL} \propto 
 \sum_{i=1}^3 m_i \tilde{U}_{ei}^2 \mathcal{M}^{0\nu\beta\beta}(0)=
 \sum_{i=1}^3 \left(U_{pmns}mU_{pmns}^T\right)_{ee} \mathcal{M}^{0\nu\beta\beta}(0),
\label{ASM}
\ee
which is the standard light neutrino contribution.

\item
$\mathbf{W_R-W_R}$ \textbf{channel}. The amplitude of the bottom left diagram of Fig.~\ref{feyn}, in which two $W_R$ are involved,
is given by~\cite{Vergados:1985pq}
\be
A_{RR} \propto \left(\frac{M^2_{W_L}}{M^2_{W_R}}+\xi\right)^2\left[\sum_{i=1}^3 m_i A_{ei}^{*\,2} \mathcal{M}^{0\nu\beta\beta}(0)+
\sum_{i=1}^3M_i V_{ei}^{*\,2} \mathcal{M}^{0\nu\beta\beta}(M_i)\right].\nonumber
\ee
Using Eq.~(\ref{U}) in the above equation we obtain:

\be
A_{RR}\propto \left(\frac{M^2_{W_L}}{M^2_{W_R}}+\xi\right)^2\left[\sum_{i=1}^3M_i V_{ei}^{*\,2}
\mathcal{M}^{0\nu\beta\beta}(M_i)-\left(\theta^T
U^*_{pmns} m U_{pmns}^\dagger \theta\right)_{ee}
\mathcal{M}^{0\nu\beta\beta}(0)\right],
\label{ARR}
\ee
Clearly, the second term can be neglected in comparison with the standard contribution
due to the double suppression coming from $\left(\frac{M^2_{W_L}}{M^2_{W_R}}+\xi\right)^2$
and the active-heavy mixing, at least $|\theta_{\alpha i}|^2\lesssim 10^{-2}$~\cite{Antusch:2006vwa,FernandezMartinez:2007ms,Atre:2009rg}.
The first term, however, can not be neglected, i.e,
\be
A_{RR}\propto\sum_{i=1}^3M_i V_{ei}^{*\,2}  \left(\frac{M^2_{W_L}}{M^2_{W_R}}+\xi\right)^2
\mathcal{M}^{0\nu\beta\beta}(M_i).
\label{ARRh}
\ee

\item
$\mathbf{W_L-W_R}$ \textbf{channel}. For the diagram in the top right of Fig.~\ref{feyn}, in which $W_L$ and $W_R$ are exchanged, the
amplitude is given by

\be
A_{LR} \propto \left(\xi+\eta\frac{M^2_{W_L}}{M^2_{W_R}}\right)\langle p\rangle\left[\sum_{i=1}^3 A_{ei}\tilde{U}_{ei}^* \mathcal{M}^{0\nu\beta\beta}(0)+
\sum_{i=1}^3 V_{ei} B_{ei}^* \mathcal{M}^{0\nu\beta\beta}(M_i)\right],
\label{ALR0}
\ee
where $\eta\approx 10^{-2}$~\cite{Doi:1985dx,Muto:1989cd,Caurier:1996zz,Hirsch:1996qw,Pantis:1996py,Suhonen:1998ck}
\footnote{Notice that the first and second terms in Eq.~(\ref{ALR0}) correspond to the 
usually called $\eta$ and $\lambda$ mechanisms respectively}. Taking into account that $U$ 
is unitary, we have:
\be
A_{LR} \propto
\left(\xi+\eta\frac{M^2_{W_L}}{M^2_{W_R}}\right)\langle p\rangle\sum_{i=1}^3 V_{ei} B^*_{ei}( 
\mathcal{M}^{0\nu\beta\beta}(M_i)-
\mathcal{M}^{0\nu\beta\beta}(0)).\label{ALR}
\ee
and since $\mathcal{M}^{0\nu\beta\beta}(M_i)/\mathcal{M}^{0\nu\beta\beta}(0)\ll1$, Eq.~(\ref{ALR}) becomes
\be
A_{LR} \propto
\left(\xi+\eta\frac{M^2_{W_L}}{M^2_{W_R}}\right)\langle p\rangle\sum_{i=1}^3 A_{ei} \tilde{U}^*_{ei}\mathcal{M}^{0\nu\beta\beta}(0)),
\label{ALRh}
\ee
which implies that the light neutrino mediated contribution of the $W_L-W_R$ channel is again dominant
over the heavy neutrino exchange.

\item The amplitude corresponding to the scalar triplet $\Delta_L$ exchange
(bottom right in Fig.~\ref{feyn} with $W_L$ and $\Delta_L$) is suppressed with the
factor

\be
\frac{(M_L)_{ee}}{\sum_i \tilde{U}_{ei}m_i}\frac{\langle p^2\rangle}{M_{\Delta_L}^2}=\frac{(M_L)_{ee}}{\left(M_L-m_DM_R^{-1}m_D^T\right)_{ee}}\frac{\langle p^2\rangle}{M_{\Delta_L}^2},
\ee
with respect to the standard contribution given in Eq.~(\ref{ASM}). The suppression factor is at 
least $ \langle p^2\rangle/M_{\Delta_L}^2\ll1$ if no fine tuned
cancellation between the two terms in the light neutrino contribution is invoked, i.e., the contribution
of this channel is negligible. For the corresponding ``right-handed'' version of the diagram the
situation is slightly different and the suppression factor now reads

\be
\frac{M_{W_L}^4}{M_{W_R}^4}\frac{(M_R)_{ee}}{\left(M_L-m_DM_R^{-1}m_D^T\right)_{ee}}\frac{\langle p^2\rangle}{M_{\Delta_R}^2}
\ee
It seems that for small enough values of
$\left(M_L-m_DM_R^{-1}m_D^T\right)_{ee} = \sum_i \left[\left(U_{pmns}\right)_{ei}\right]^2 m_i$,
this contribution could be larger than the standard one. However, it is not very easy to achieve
a measurable  $\Delta_R$ contribution, at the reach of the sensitivity of the next-to-next of $0\nu\beta\beta$
decay experiments ($m_{\beta\beta}\sim10^{-2}$ eV). Indeed, the corresponding amplitude is given by
\be
A_{\Delta_R} \propto (M_R)_{ee}\,
\mathcal{M}_{\Delta}^{0\nu\beta\beta}(M_{\Delta_R})\approx
 \frac{M_{W_L}^4}{M_{W_R}^4}\frac{\langle p^2\rangle}{v_R}\frac{\left(Y_\Delta\right)_{ee}}{2\rho}\mathcal{M}^{0\nu\beta\beta}(0),
\label{AD}
\ee
where we have used $M_{\Delta_R}^2\approx 2\rho v_R^2$ and $\mathcal{L} \supset \rho \mbox{Tr}
\left( \Delta_R \Delta_R^{\dagger} \Delta_R \Delta_R^{\dagger}\right)+ Y_\Delta \bar{L_R^c}
\Delta_R L_R$.\footnote{We refer readers to Ref.~\cite{Mohapatra:1980yp}
for more details}
The only possibility of having a phenomenologically relevant contribution is to
saturate the bounds on $v_R$ ($M_{W_L}^4/M_{W_R}^4<10^{-6}$~\cite{Aad:2012ej} and $M_{W_R}\approx v_Rg/\sqrt{2}$) having at the same time
$Y_\Delta\gg\rho$, which is not very feasible since a small value of $\rho$ would render $\Delta_R$ too light,
contradicting the experimental bound, $m_{\Delta_R}>320$ GeV~\cite{ATLAS:2012hi}. We will thus neglect
this contribution.

\end{itemize}

We have shown that only the contributions coming from the light neutrino exchange can have
a significant impact in the $0\nu\beta\beta$ decay rate, with the exception of the
channel mediated by two $W_R$ gauge bosons in which the heavy neutrino exchange dominates.
In summary, the phenomenologically relevant contributions to the $0\nu\beta\beta$ decay
rate can be recast as

\bea
A_{total}&\propto& \left[ c_{LL}\sum_{i=1}^3 \left[\left(U_{pmns}\right)_{ei}\right]^2 m_i +
c_{RR}\left(\frac{M^2_{W_L}}{M^2_{W_R}}+\xi\right)^2\sum_{i=1}^3M_i V_{ei}^{*\,2} 
\frac{\mathcal{M}^{0\nu\beta\beta}(M_i)}{\mathcal{M}^{0\nu\beta\beta}(0)}\right.
\nonumber\\
&-&\left.  c_{LR}\,\theta^*_{e1}\left(\xi+\eta\frac{M^2_{W_L}}{M^2_{W_R}}\right)\langle p\rangle\right]
\mathcal{M}^{0\nu\beta\beta}(0)
\nonumber\\
&\equiv& m_{\beta\beta}\,{\mathcal{M}^{0\nu\beta\beta}(0)},
\eea
where we have made use of Eq.~(\ref{U}) and $c_{LL}$, $c_{LR}$ and $c_{RR}$
are coefficients which take into account the different chirality of the outgoing electrons. At this point
we can make an estimation of the NMEs associated to the heavy neutrino exchange, $\mathcal{M}^{0\nu\beta\beta}(M_i)$, to understand how
relevant the remaining NP contributions are. The effective mass becomes,
\be
|m_{\beta\beta}|^2= \left\lvert\left(\frac{v_L}{v_R}M_R-\theta M_R \theta^T\right)_{ee}\right\rvert^2
+\left\lvert\left(\xi+\eta\frac{M^2_{W_L}}{M^2_{W_R}}\right)\langle p\rangle\theta_{e1}\right\rvert^2
+\left\lvert\left(\frac{M^2_{W_L}}{M^2_{W_R}}+\xi\right)^2\langle p^2\rangle\left[\left(M_R\right)^{-1}\right]_{ee}
\right\rvert^2
\label{mbb}
\ee
where we have neglected the suppressed interference terms between the different chirality contributions~\cite{Halprin:1983ez}.

In the rest of this section, we will first study the bounds that can be extracted from the $0\nu\beta\beta$ decay experiments if
one assumes that the three contributions listed in Eq.~(\ref{mbb}) are completely independent. After that, we will study
the region of the parameter space in which a NP signal in the future $0\nu\beta\beta$ decay experiments can be expected when
the correlations among the different contributions are not ignored.

\begin{figure}
\begin{tabular}{cc}
\includegraphics[width=0.5\textwidth,angle=0]{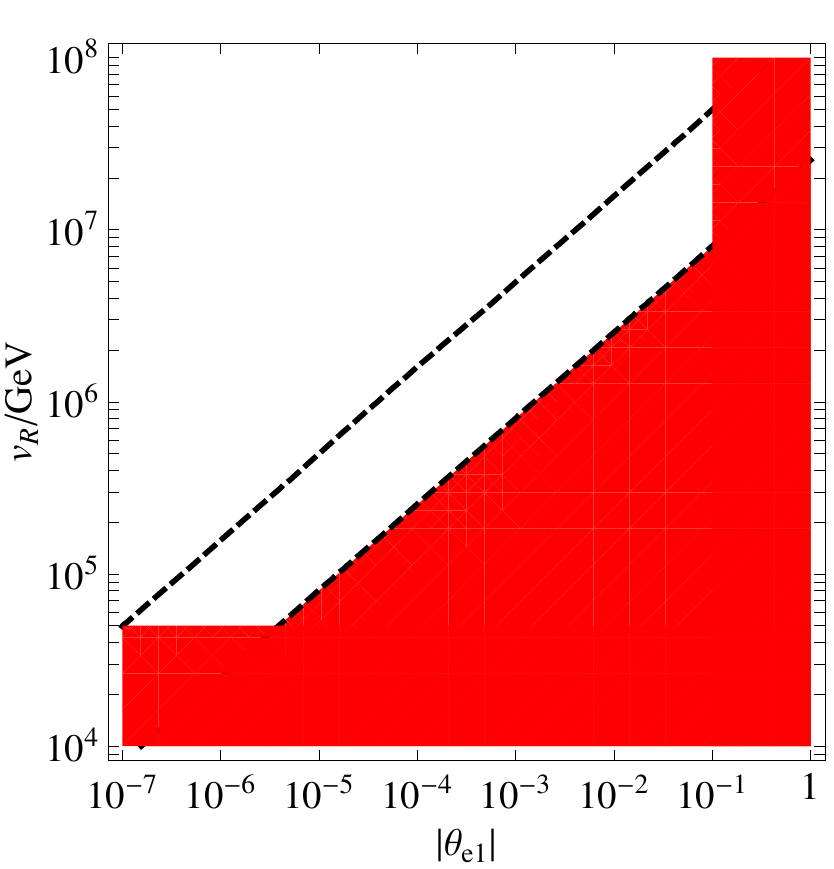} &
\includegraphics[width=0.5\textwidth,angle=0]{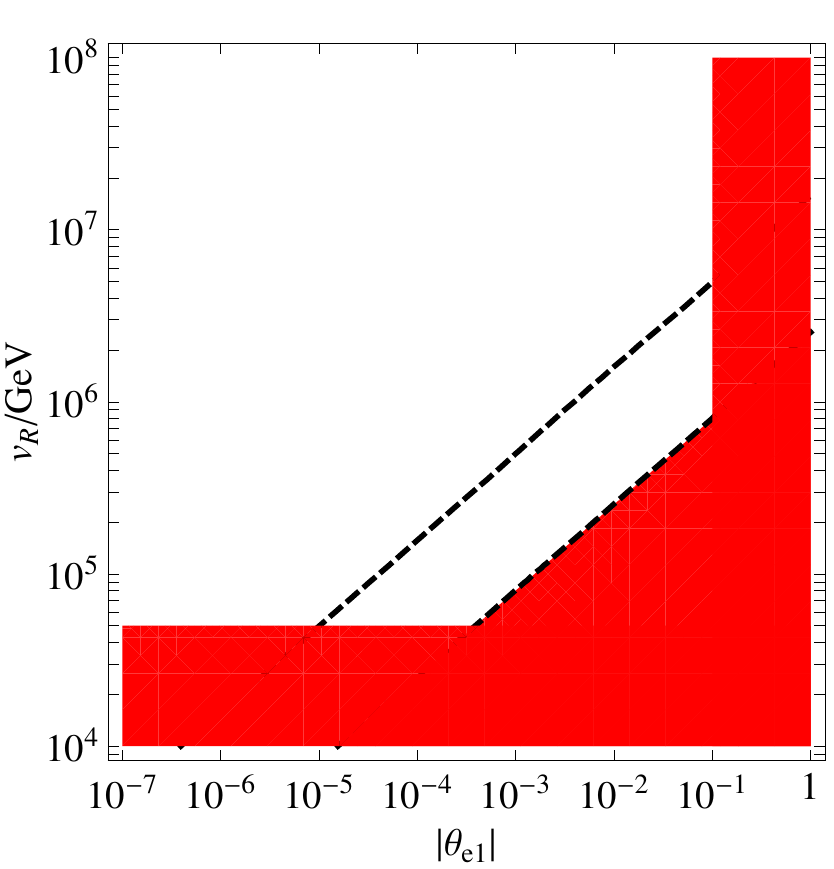}
\end{tabular}
\caption{\label{figLR}
 \textbf{Heavy regime.} The shaded region represents the values of $v_R$ and $\lvert\theta_{e1}\rvert$ ruled out by the present
experimental bound on the $0\nu\beta\beta$ decay rate mediated by the $W_L-W_R$ channel (neglecting the standard and the $W_R-W_R$
contributions) and the bounds on $M_{W_R}$~\cite{Aad:2012ej} and non-unitarity. The future $0\nu\beta\beta$ decay
sensitivity, when the standard light neutrino and the $W_R-W_R$ contributions are not included, is given by the region 
between the red dashed lines. The mixing $\xi$ has been fixed to $M_{W_L}^2/M_{W_R}^2$ (zero) in the left (right) panel. }
\end{figure}

In Fig.~\ref{figLR} we show the constraints on $v_R$
(recall that $M_{W_R}=gv_R/\sqrt{2}$) and the mixing between $\nu_{eL}$ and the lightest heavy neutrino,
$\lvert\theta_{e1}\rvert$, extracted from $0\nu\beta\beta$ decay experiments when only the contribution from the $W_L-W_R$
channel (second term in Eq.~(\ref{mbb})) is taken into account. In the left panel the mixing $\xi$ saturates the 
theoretical bound ($\xi=M_{W_L}^2/M_{W_R}^2$) while in the right panel $\xi$ is neglected. The shaded region is ruled out by the present constraint, $\lvert m_{\beta\beta}\rvert< 0.38$ eV~\cite{Auger:2012ar}, 
while the region between the red dashed lines corresponds to the sensitivity of the next-to-next generation of experiments, $10^{-2}$ eV $<\lvert m_{\beta\beta}\rvert< 0.38$ eV.

\begin{figure}
\begin{tabular}{cc}
\includegraphics[width=0.5\textwidth,angle=0]{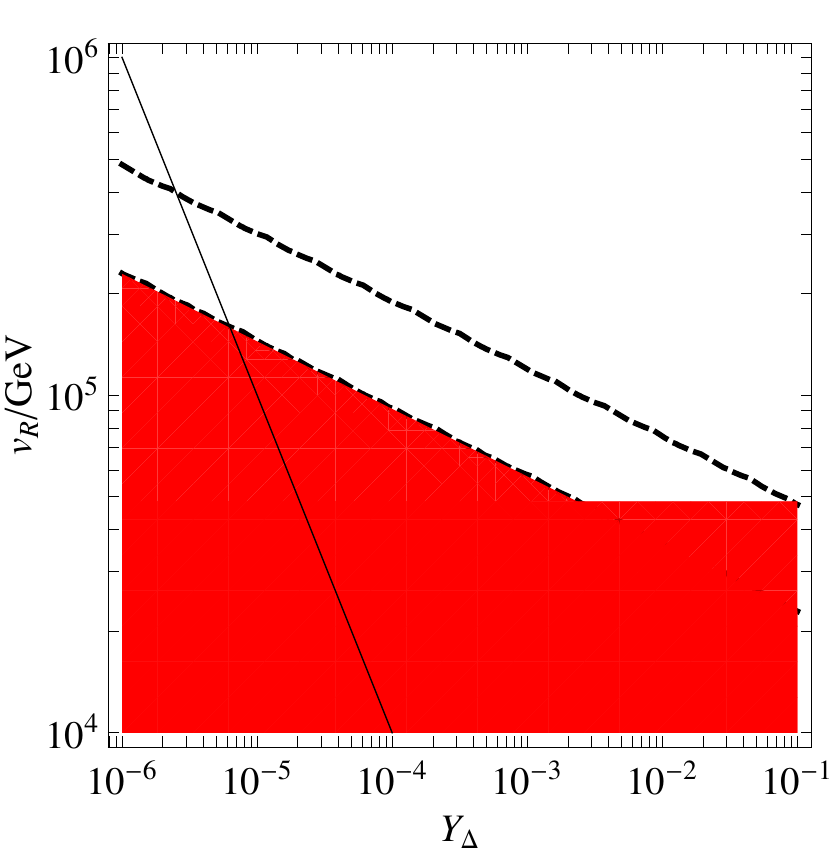} &
\includegraphics[width=0.5\textwidth,angle=0]{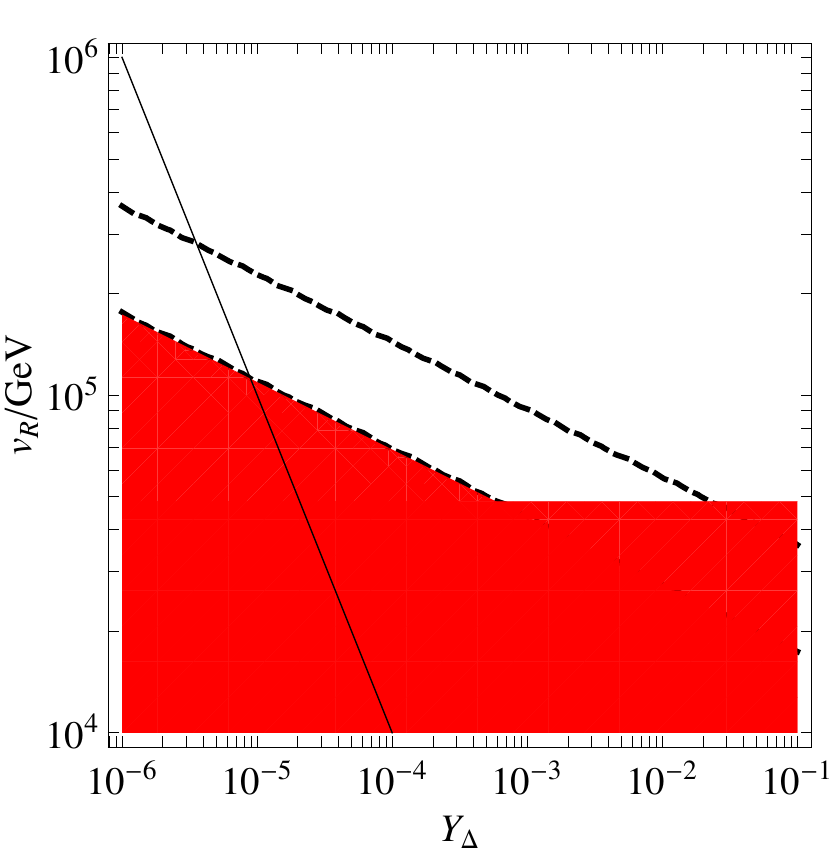}
\end{tabular}
\caption{\label{figRR}
\textbf{Heavy regime.} The shaded region represents the values of $v_R$ and $\left(Y_\Delta\right)_{ee}$ ruled out by the present
experimental bound on the $0\nu\beta\beta$ decay rate mediated by the $W_R-W_R$ channel (neglecting the
standard and the $W_L-W_R$
contributions) and the bounds on $M_{W_R}$~\cite{Aad:2012ej}. The future $0\nu\beta\beta$ decay
sensitivity, when the standard light neutrino and the $W_L-W_R$ contributions are not included, is given
by the region between the red dashed lines. The black line corresponds to $\left(M_R\right)_{ee}=1$ GeV. 
The mixing $\xi$ has been fixed to $M_{W_L}^2/M_{W_R}^2$ (zero) in the left (right) panel.}
\end{figure}

Fig.~\ref{figRR} is analogous to Fig.~\ref{figLR}, but this time we show the present bound on $v_R$ and $\left(Y_\Delta\right)_{ee}$
when only the contribution from the $W_R-W_R$ channel is considered, i.e., only the third term of
Eq.~(\ref{mbb}) is included in the analysis. The future sensitivity is shown as well.

The caveat for Figs.~\ref{figLR} and \ref{figRR} is that we switch on the NP contributions one at a time
without considering the correlation between them and that of the light neutrinos. This is specially problematic
if one tries to find the future sensitivity to the parameters of the model. For example, 
from Fig.~\ref{figLR} one would conclude that $v_R$ can be probed in the region $50$ TeV $\lesssim v_R\lesssim 4\cdot 10^{4}$ 
TeV while from Fig.~\ref{figRR} the conclusion would be different, probing $50$ TeV $\lesssim v_R\lesssim 500$ TeV. In this context, the
following two questions arise. First, is the standard light neutrino contribution significant for those inputs of the
parameters? Can those NP contributions really dominate over the standard one? And second, if yes, for what
region of the parameter space? The $0\nu\beta\beta$ decay phenomenology is sometimes analyzed taking into account
the different contributions one by one, this is, neglecting the rest of the contributions and the
correlations induced by the neutrino mass generation mechanism. In this work, we simultaneously include all the
relevant contributions in the analysis and emphasize how the correlation plays a vital role in order to answer
the previous questions.

\begin{figure}
\begin{tabular}{cc}
  \includegraphics[width=0.5\textwidth,angle=0]{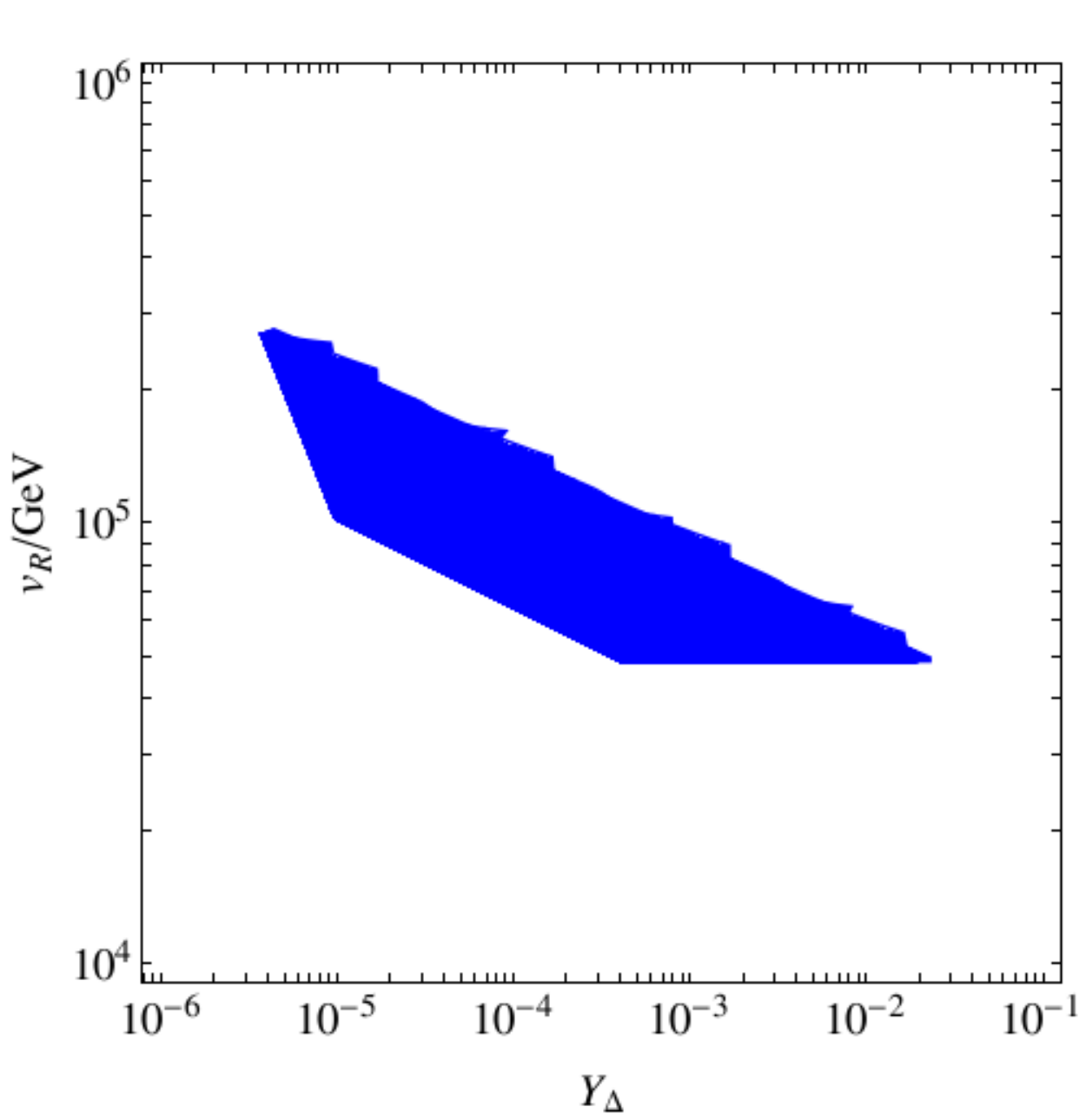} &
    \includegraphics[width=0.5\textwidth,angle=0]{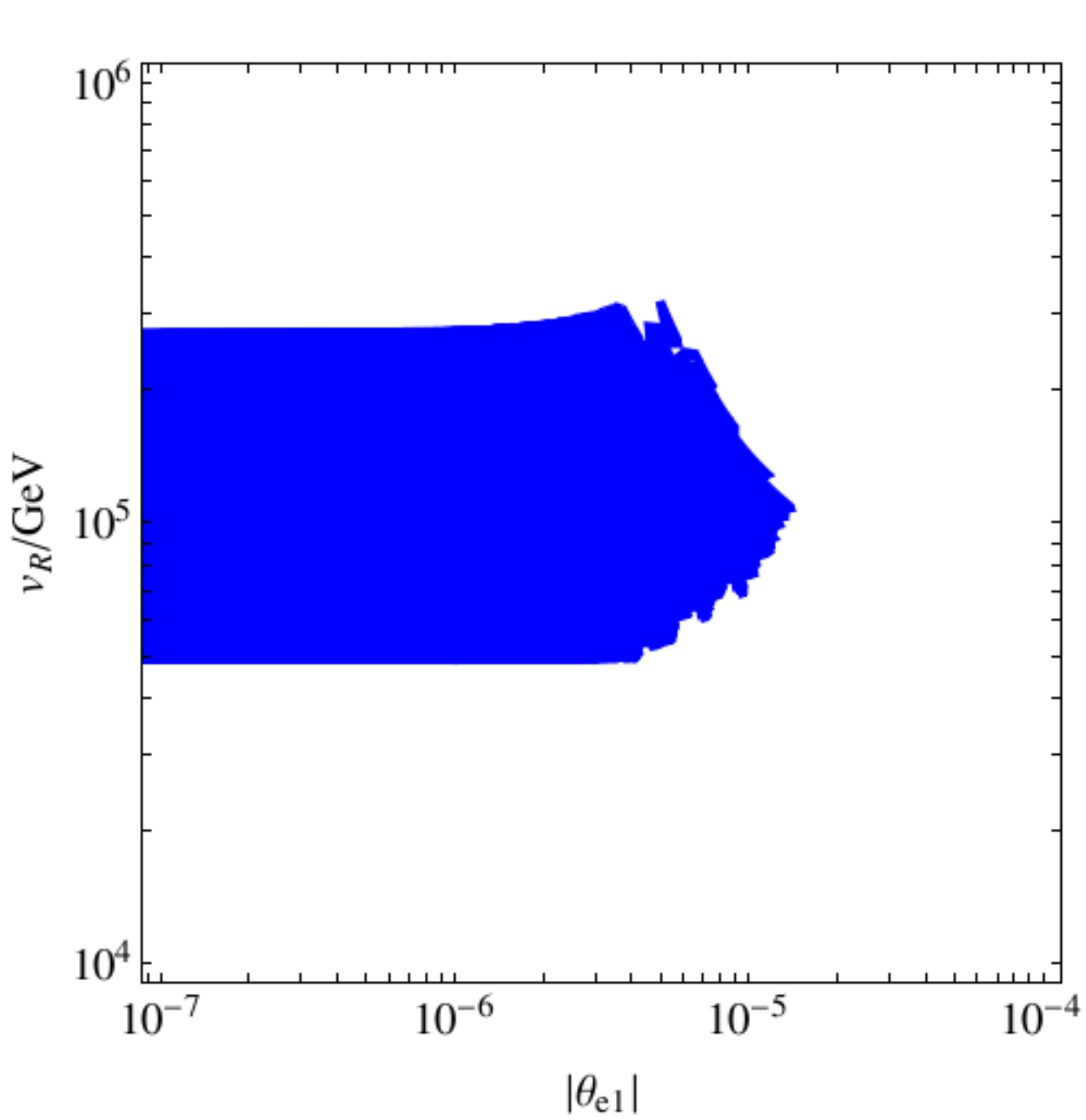}
 \end{tabular}
\begin{tabular}{cc}
  \includegraphics[width=0.5\textwidth,angle=0]{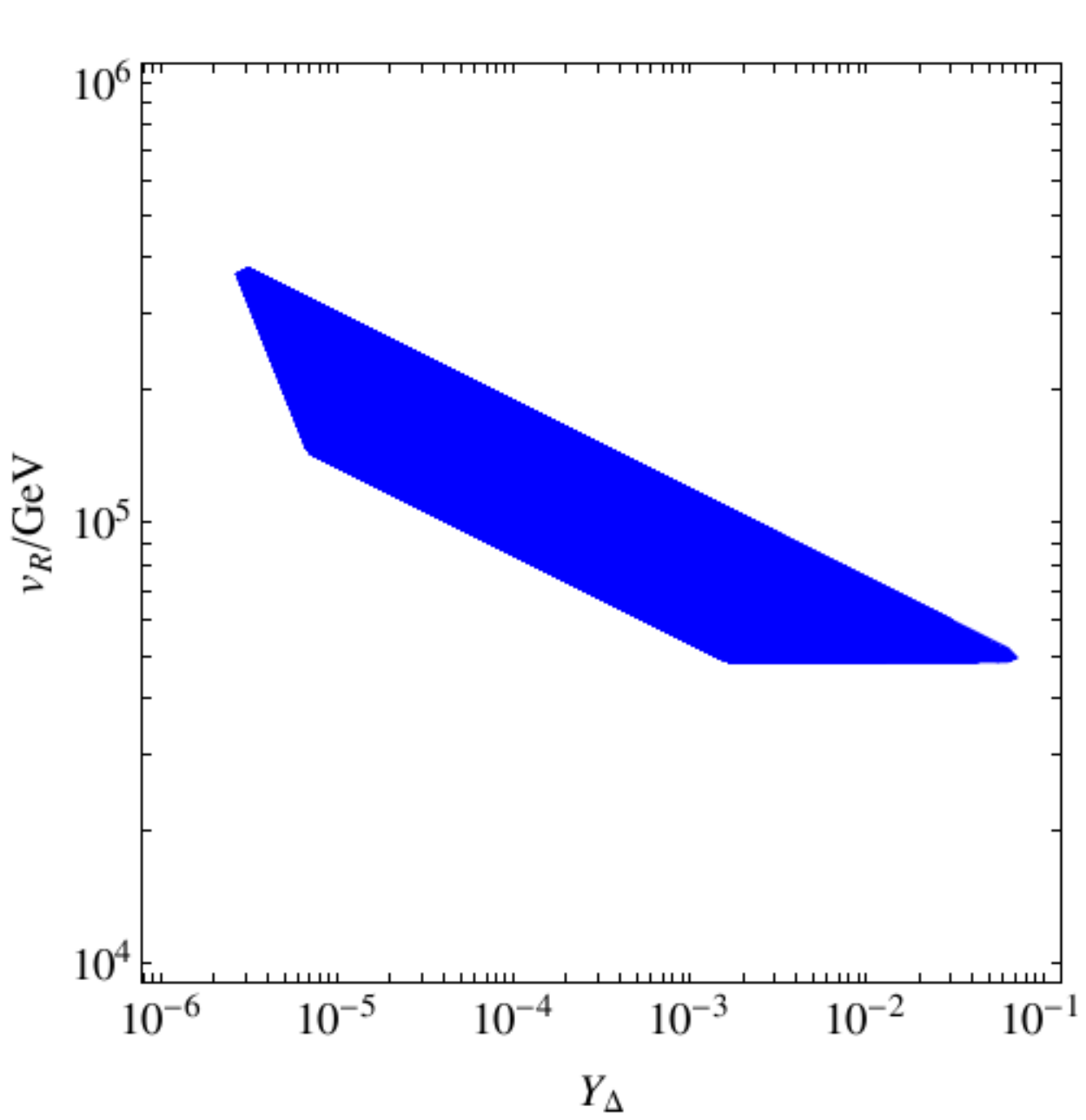} & 
    \includegraphics[width=0.5\textwidth,angle=0]{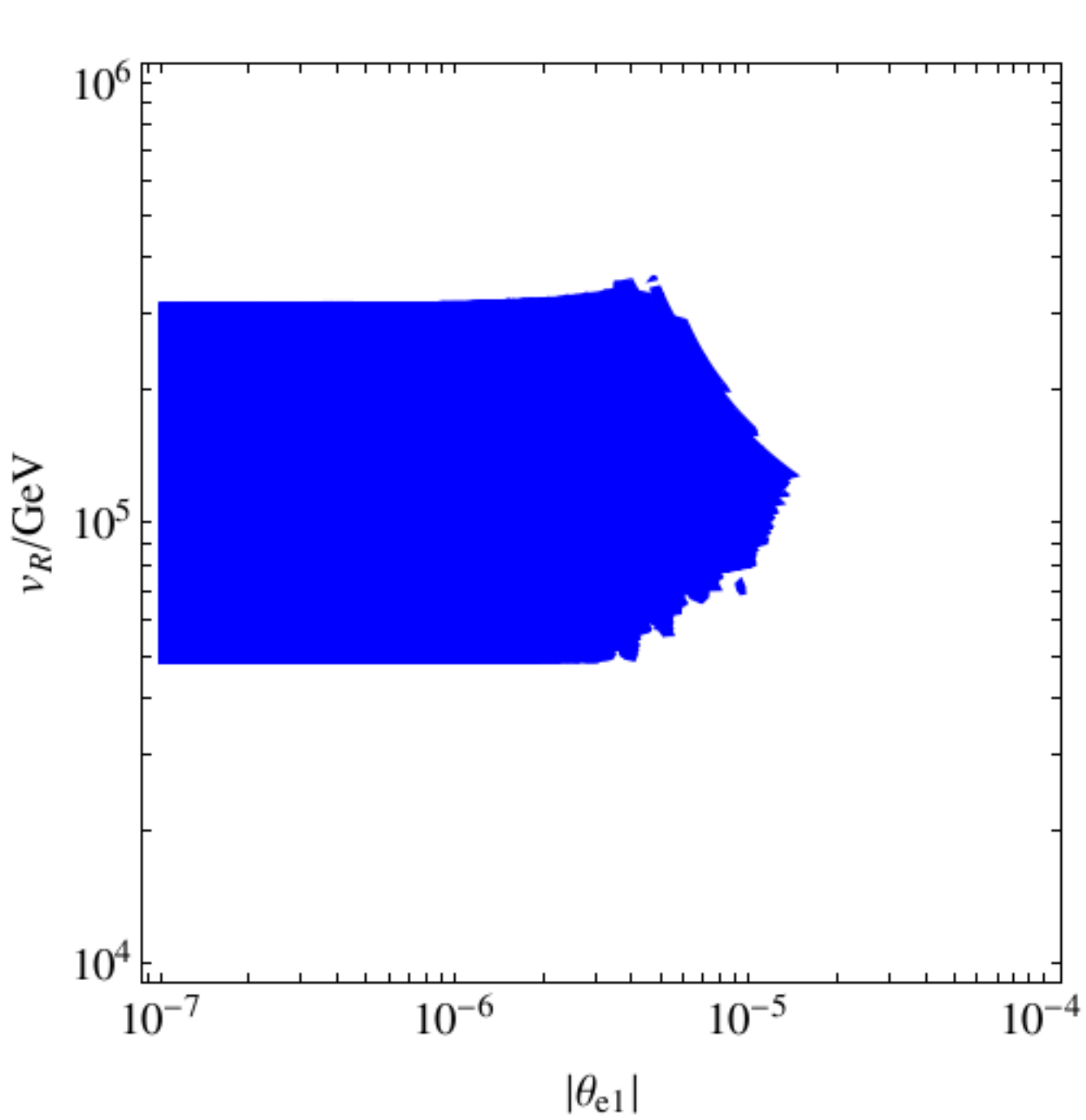}
 \end{tabular}
\caption{\label{figheavy}
\textbf{Heavy regime.} The shaded region represents the future sensitivity of the next-to-next generation of $0\nu\beta\beta$ decay
experiments to $v_R$ and $\left(Y_\Delta\right)_{ee}$ (left panel) and $v_R$ and $\lvert \theta \rvert$ (right
panel) when the decay rate is dominated by the NP contribution for $\xi=0$ (upper panels) and 
$\xi=M_{W_L}^2/M_{W_R}^2$ (lower panels). In the calculation all the relevant contributions to the process have been included at the same
time in the analysis and the present bound on $M_{W_R}$~\cite{Aad:2012ej} and the active-heavy neutrino mixing $\theta$
is respected. A hierarchical heavy neutrino spectrum has been considered.}
\end{figure}

In Fig.~\ref{figheavy} we show the sensitivity of the next-to-next generation of $0\nu\beta\beta$ decay
experiments ($10^{-2}$ eV $<\lvert m_{\beta\beta}\rvert< 0.38$ eV) to the parameters of the model
by including all the relevant contributions and requiring the NP contribution to the $0\nu\beta\beta$ decay
rate (second and third term in Eq.~(\ref{mbb})) to be at least 10 times larger than the standard contribution
(first term of Eq.~(\ref{mbb})). The allowed region is projected onto the
$v_R$-$\left(Y_\Delta\right)_{ee}$ plane (left panel) and the $v_R$-$\lvert \theta_{e1} \rvert$
plane (right panel). The mixing has been neglected in the upper panels while in the 
lower panels is fixed to its maximum value $\xi=M_{W_L}^2/M_{W_R}^2$. The experimental constraints on the $W_R$ mass~\cite{Aad:2012ej} and the active-heavy mixing~\cite{Antusch:2006vwa,FernandezMartinez:2007ms,Atre:2009rg}
have been also included. We have assumed that the heavy
neutrino spectrum is hierarchical ($M_1\ll M_2,M_3$). We confirm that a dominant contribution in the 
left-right symmetric model coming from NP channels is still
possible for the window $50$ TeV$\lesssim v_R \lesssim 300$ TeV ($50$ TeV$\lesssim v_R \lesssim 400$ TeV) 
if the trilinear coupling and the mixing are small enough,  
$3\cdot10^{-6} \lesssim\left(Y_\Delta\right)_{ee}\lesssim 3\cdot 10^{-2}$ 
($3\cdot10^{-6} \lesssim\left(Y_\Delta\right)_{ee}\lesssim 8\cdot 10^{-2}$)
and $\lvert \theta_{e1} \rvert\lesssim 2\cdot 10^{-5}$ respectively, for $\xi=0$ (maximal mixing $\xi=M^2_{W_L}/M^2_{W_R}$). This corresponds to a range of heavy neutrino masses
from GeV to TeV. Comparing the upper and lower panels we can conclude that including the mixing 
in the analysis has some impact in the results but it is not very significant.

Comparing Fig.~\ref{figheavy} with Fig.~\ref{figLR}, where only the $W_L-W_R$ contribution is included,
we see that the region of the parameter space which can be experimentally probed shrinks when all the
contributions are included at once. From Fig.~\ref{figLR}, one could conclude that a NP signal from
the $W_L-W_R$ channel is possible for very large values of $v_R$ up to $\sim10^{4}$ TeV
($M_{W_R}\sim 500$ TeV). However, such a large value of $v_R$ requires a quite large mixing
$\theta$ since $v_R$ suppresses the $W_L-W_R$ contribution (second term of Eq.~(\ref{mbb}))
which makes the light neutrino contribution to the $0\nu\beta\beta$ decay rate larger than the
present bound. Namely, due to the correlation, such a large values of $v_R$ and $\theta_{e1}$ are
 ruled out and the $W_L-W_R$ channel can not give a dominant contribution to the process.
 On the other hand, even for values of $v_R$ close to the present bound, the $W_L-W_R$ contribution 
 is of the same order of the subleading light active neutrino one, while the $W_R-W_R$ contribution becomes larger than that of the $W_L-W_R$ channel or even above the present
experimental bound. Indeed, we have checked numerically that,
once the correlations are taken into account, a NP signal can be expected mainly from the $W_R-W_R$ channel
as one can anticipate from the fact that the NP signal regions in Fig.~\ref{figRR} and Fig.~\ref{figheavy} (left) overlap.
Note that the $W_L-W_R$ channel could only dominate the decay rate if a cancellation in the light 
neutrino contribution takes place, a scenario not explored in this work.

\subsection{Light regime}
\label{lightregime}

If the RH neutrinos are lighter than the $0\nu\beta\beta$ decay scale, $\mathcal{O}\left(100\right)$ MeV,
the picture significantly changes with respect to the heavy scenario studied above. Eqs.~(\ref{ALL}),
(\ref{ARRh}) and (\ref{ALR})
remain valid but the NMEs associated with the ``heavy'' neutrino exchange are not suppressed compared to the light
neutrino mediated ones. In fact, for $M_i<1$ MeV, we have in a very good approximation
$\mathcal{M}^{0\nu\beta\beta}(0)=\mathcal{M}^{0\nu\beta\beta}(M_i)$. This yields a cancellation within the second term of
Eqs.~(\ref{ALL}) and the corresponding amplitude is then given by

\be
A_{LL} \propto (M_L)_{ee}\, \mathcal{M}^{0\nu\beta\beta}(0)
=\frac{v_L}{v_R} (M_R)_{ee}\,\mathcal{M}^{0\nu\beta\beta}(0),
\label{ALLl}
\ee
while Eq.~(\ref{ARRh}) becomes
\be
A_{RR} \propto 
\left(\frac{M^2_{W_L}}{M^2_{W_R}}+\xi\right)^2(M^*_R)_{ee}\,
\mathcal{M}^{0\nu\beta\beta}(0),\label{ARRl}
\ee
where we have used Eq.~(\ref{constraints3}). The $W_L-W_R$ contribution, given by Eq.~(\ref{ALR}), vanishes
due to the unitarity of the $6\times6$ neutrino mixing matrix $U$. Therefore, from Eqs.~(\ref{ALLl})-(\ref{ARRl}), the expression for the effective mass $m_{\beta\beta}$
when the RH neutrinos are lighter than $1$ MeV becomes:

\bea
|m_{\beta\beta}|^2&=& \left\lvert\frac{v_L}{v_R}\left(M_R\right)_{ee}\right\rvert^2
+\left\lvert\,\left(M^*_R\right)_{ee}\left(\frac{M^2_{W_L}}{M^2_{W_R}}+\xi\right)^2\right\rvert^2
\nonumber\\
&=&
\left\lvert v_L\left(Y_\Delta\right)_{ee}\right\rvert^2
+\left\lvert\,v_R\left(Y^*_\Delta\right)_{ee}\left(\frac{M^2_{W_L}}{M^2_{W_R}}+\xi\right)^2\right\rvert^2.
\label{mbb2}
\eea
We conclude that in this regime the $0\nu\beta\beta$ decay can be completely
attributed to the $W_R-W_R$ channel contribution,
if $v_L/v_R\ll \left(M_{W_L}/M_{W_R}\right)^4$.\footnote{Note that the mass
of $M_{W_L}$ mainly comes from the SM Higgs vev, not from the $\Delta_L$ vev,
$v_L$. As a consequence, $v_L$ could be very small. Furthermore, a small $v_L$ is in better agreement with
the $\rho(\equiv M^2_{W_L}/M^2_{Z_L}\cos^2\theta_W)$ parameter constraints.}

If $v_L/v_R\gg \left(M_{W_L}/M_{W_R}\right)^4$, the $W_L-W_L$ channel dominates the decay rate. This does
not mean that the standard contribution (that mediated by the light active neutrinos) always dominates since
the RH neutrino exchange can also contribute in this channel. Indeed,

\be
A_{LL} \propto \left(U_{pmns}\,m\,U_{pmns}^T +\theta M_R \theta^T \right)_{ee} \, \mathcal{M}^{0\nu\beta\beta}(0)=
\frac{v_L}{v_R}(M_R)_{ee} \mathcal{M}^{0\nu\beta\beta}(0),\label{ALLl2}
\ee
Notice that, contrary to the type-I seesaw limit ($v_L\rightarrow 0$)~\cite{Blennow:2010th}, in 
this regime $A_{LL}$ does not vanish and the RH neutrinos (second term in the equation above) 
can contribute to the process. Nevertheless, in this work we will focus on the $W_R-W_R$ and $W_L-W_R$ channels. A dominant NP
contribution from $W_L-W_L$ channel mediated by the RH neutrinos will be investigated elsewhere in more detail.

\begin{figure}
\begin{tabular}{ccc}
  \includegraphics[width=0.33\textwidth,angle=0]{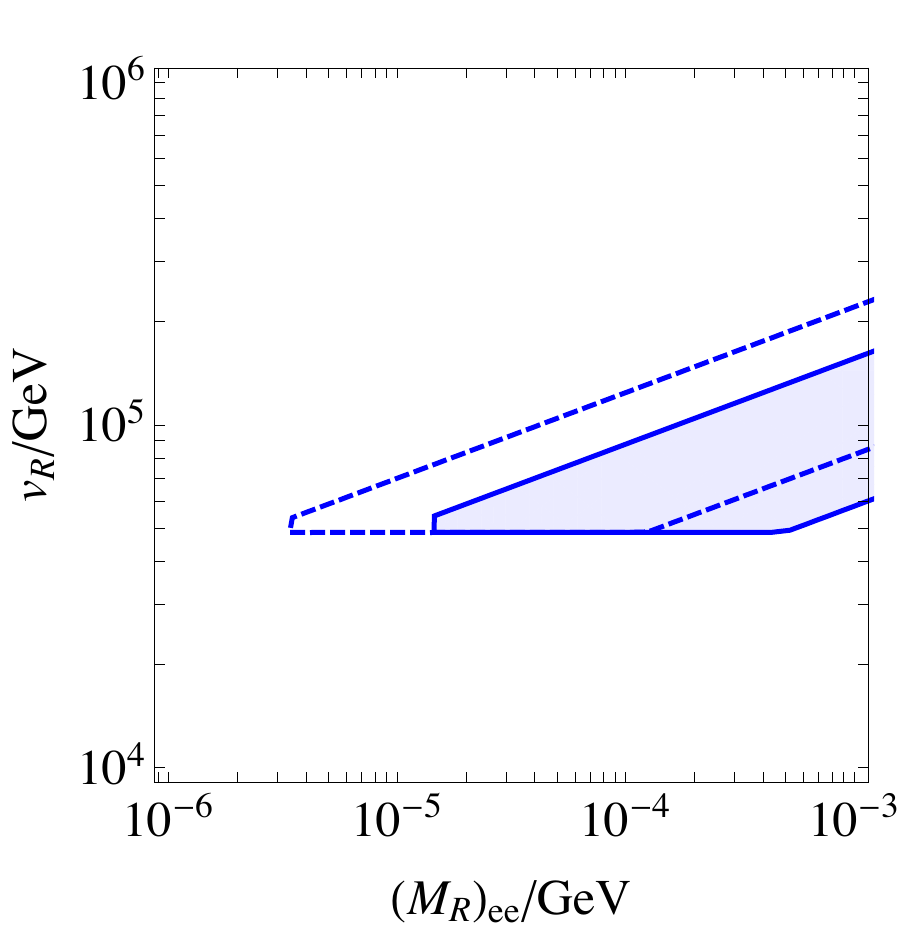} &
    \includegraphics[width=0.33\textwidth,angle=0]{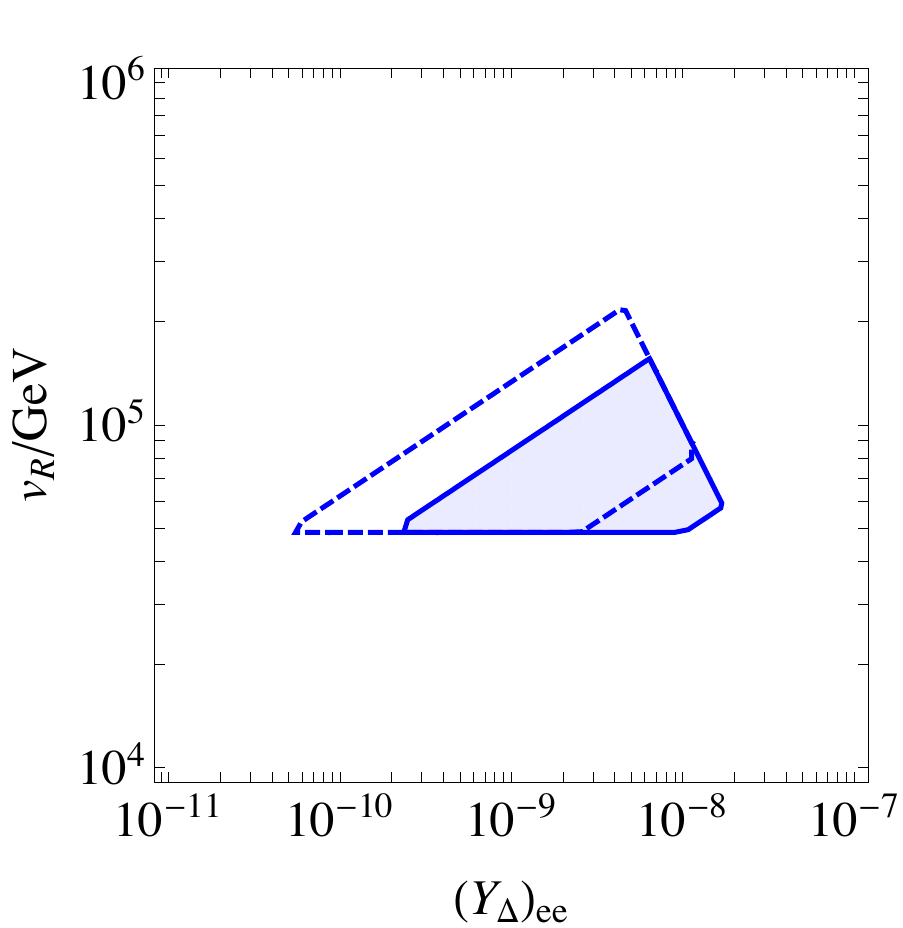} &
 \includegraphics[width=0.33\textwidth,angle=0]{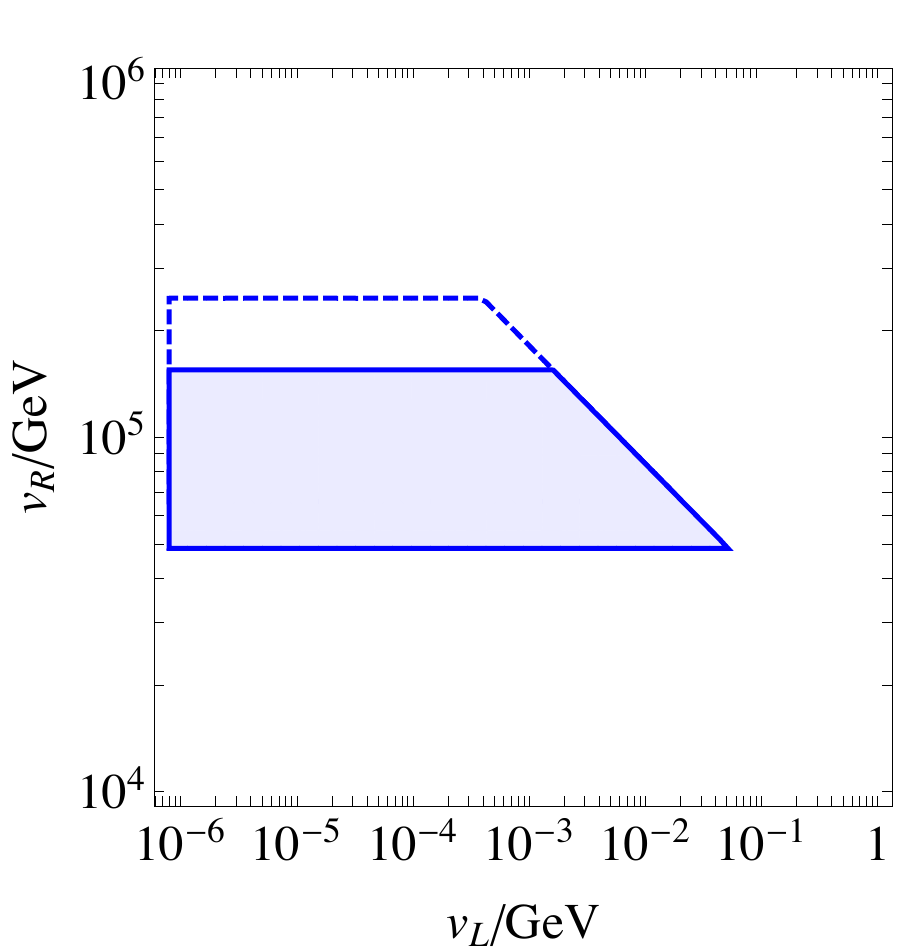}
 \end{tabular}
\caption{\label{figlight1}
\textbf{Light regime.} The region inside the solid (dashed) lines represents the future sensitivity 
of the next-to-next generation of $0\nu\beta\beta$ decay experiments when the decay rate is dominated 
by the $W_R-W_R$ contribution for $\xi=0$ ($\xi=M_{W_L}^2/M_{W_R}^2$), projected onto the $v_R-(M_R)_{ee}$
plane (left panel), $v_R-\left(Y_\Delta\right)_{ee}$ (central panel) and $v_R-v_L$ (right panel). In the analysis all the
relevant contributions have been simultaneously included. The bounds on $M_{W_R}$~\cite{Aad:2012ej} have been also included.}
\end{figure}

The future sensitivity of the next-to-next generation of $0\nu\beta\beta$ decay experiments ($10^{-2}$ eV
$<\lvert m_{\beta\beta}\rvert< 0.54$ eV) to the parameters of the model when the $W_R-W_R$ contribution (second term of Eq.~(\ref{mbb2})) 
is at least 10 times larger than that from the $W_L-W_L$ channel (first term of Eq.~(\ref{mbb2})) 
is given in Fig.~\ref{figlight1}. The allowed region of the parameter space is projected this time onto
the $v_R-\left(M_R\right)_{ee}$ plane (left panel), $v_R-\left(Y_{\Delta}\right)_{ee}$ (central panel) 
and $v_R-v_L$ (right panel) for $\xi=0$ (solid line) and $\xi=M_{W_L}^2/M_{W_R}^2$ (dashed line).
The bounds on the $W_R$ mass~\cite{Aad:2012ej} have been also included.

Fig.~\ref{figlight1} (left panel) shows that a NP dominated $0\nu\beta\beta$ decay signal can
be expected for $\left(M_R\right)_{ee}\sim15$ KeV$-1$ MeV ($\left(M_R\right)_{ee}\sim 3$ KeV$-1$ MeV) 
and $M_{W_R}\lesssim 8$ TeV ($M_{W_R}\lesssim 12$ TeV) for $\xi=0$ ($\xi=M_{W_L}^2/M_{W_R}^2$). This
means that in this regime the future sensitivity to $M_{W_R}$ is around a factor $2$ weaker than in
the heavy regime ($M_{W_R} \lesssim 15-20$ TeV depending on the value of $\xi$). In both regions the sensitivity is driven 
by the $W_R-W_R$ channel. In order to have a dominant $W_R-W_R$ contribution, the $W_L-W_L$ one should be
of course depleted and this can be achieved for small enough values of $v_L$ ($v_L\lesssim 0.07$ GeV) as
expected from Eq.~(\ref{mbb2}) and shown in Fig.~\ref{figlight1} (right panel). One may ask if
it is really feasible or natural to have RH neutrinos lighter than $1$ MeV while
$v_R$ is above the TeV. Indeed, this is perfectly possible but requires an uncomfortably small value of the
the trilinear Yukawa coupling $Y_\Delta$ since $M_i\sim Y_\Delta v_R$, as shown in Fig.~\ref{figlight1}
(left). However, the smallness of $Y_\Delta$ could be achieved adding an extra mildly broken global
symmetry to the model, as it is done in the popular inverse seesaw models with the lepton number.
In such a case these small values of $Y_\Delta$  could be considered technically natural
since $Y_\Delta=0$ would restore the global symmetry. In any case, it should be remarked that
a NP signal can only occur for $10^{-10}\lesssim \left(Y_\Delta\right)_{ee}\lesssim 10^{-8}$. Finally,
comparing the dashed and solid contours we can conclude that the impact of the mixing $\xi$ 
is not very significant in this region of the parameter space.

\subsection{Mixed Scenario}
\label{mixedregime}

There is an alternative scenario that has not been studied in the previous sections and consists of
the existence of RH neutrinos in both regimes below and above the $0\nu\beta\beta$ decay scale. In this
section we will focus on the particular case in which one of the RH neutrinos is lighter than $1$ MeV and the other
two are heavier than $1$ GeV , i.e., $M_1<1$ MeV and $M_2,M_3>1$ GeV, but the phenomenology remains
similar if two RH neutrinos are lighter than $1$ MeV.

As it occurs in the previous section, Eqs.~(\ref{ALL}), (\ref{ARRh}) and (\ref{ALR}) are also correct
in this regime, but only the NMEs associated with the $N_2$ and $N_3$ exchange are suppressed compared to
the light neutrino mediated ones. The NMEs associated with $N_1$
satisfy $\mathcal {M}^{0\nu\beta\beta}(0)=\mathcal{M}^{0\nu\beta\beta}(M_1)$. As a consequence, in this regime Eqs.~(\ref{ALL})
and (\ref{ALR}) read

\bea
A_{LL} &\propto& 
\left[ \frac{v_L}{v_R}\left(VMV^T\right)_{ee} -\sum_{i=2}^3 M_i \left(\theta V\right)_{ei}^2\right]
\mathcal{M}^{0\nu\beta\beta}(0),
\label{ALLlh}\\
A_{LR}&\propto&
-\sum_{i=2}^3 \left(\theta^* V^*\right)_{ei} V_{ei}\,\left(\xi+\eta\frac{M^2_{W_L}}{M^2_{W_R}}\right)\langle p\rangle\mathcal{M}^{0\nu\beta\beta}(0).
\label{ALRlh}
\eea
and Eq.~(\ref{ARRh}) becomes
\be
A_{RR} \propto \left( V_{e1}^{*\,2} M_1 - \sum_{i=2}^3 V_{ei}^{*\,2} \frac{ \langle p\rangle^2 }{ M_i}  \right)
\left(\frac{M^2_{W_L}}{M^2_{W_R}}+\xi\right)^2\mathcal{M}^{0\nu\beta\beta}(0),
\label{ARRlh}
\ee
where again we have used Eq.~(\ref{constraints3}) and the fact that $\mathcal{M}^{0\nu\beta\beta}(M_i)/
\mathcal{M}^{0\nu\beta\beta}(0)\ll1$ for $i=2,3$.
Therefore, in this scenario the effective mass $m_{\beta\beta}$ is given by:

\bea
|m_{\beta\beta}|^2=&& \left\lvert \frac{v_L}{v_R}\left(VMV^T\right)_{ee} -\sum_{i=2}^3 M_i \left(\theta V\right)_{ei}^2
\right\rvert^2+\left\lvert
\sum_{i=2}^3 \left(\theta V\right)^*_{ei} V_{ei}\,\left(\xi+\eta\frac{M^2_{W_L}}{M^2_{W_R}}\right)\langle p\rangle\right\rvert^2 \notag\\
+&&\left\lvert\,  \left(  V_{e1}^{*\,2} M_1 - \sum_{i=2}^3 V_{ei}^{*\,2} \frac{ \langle p\rangle^2 }{ M_i}   \right) \left(\frac{M^2_{W_L}}{M^2_{W_R}}+\xi\right)^2\right\rvert^2.
\label{mbb3}
\eea
Contrary to the light regime, in this scenario the $W_L-W_R$ contribution may be
significant. Notice that if $V=I$, the $W_L-W_R$ contribution cancels out, which means that
the RH neutrino mixing $V$ is very relevant in this region.

\begin{figure}
\begin{tabular}{ccc}
  \includegraphics[width=0.33\textwidth,angle=0]{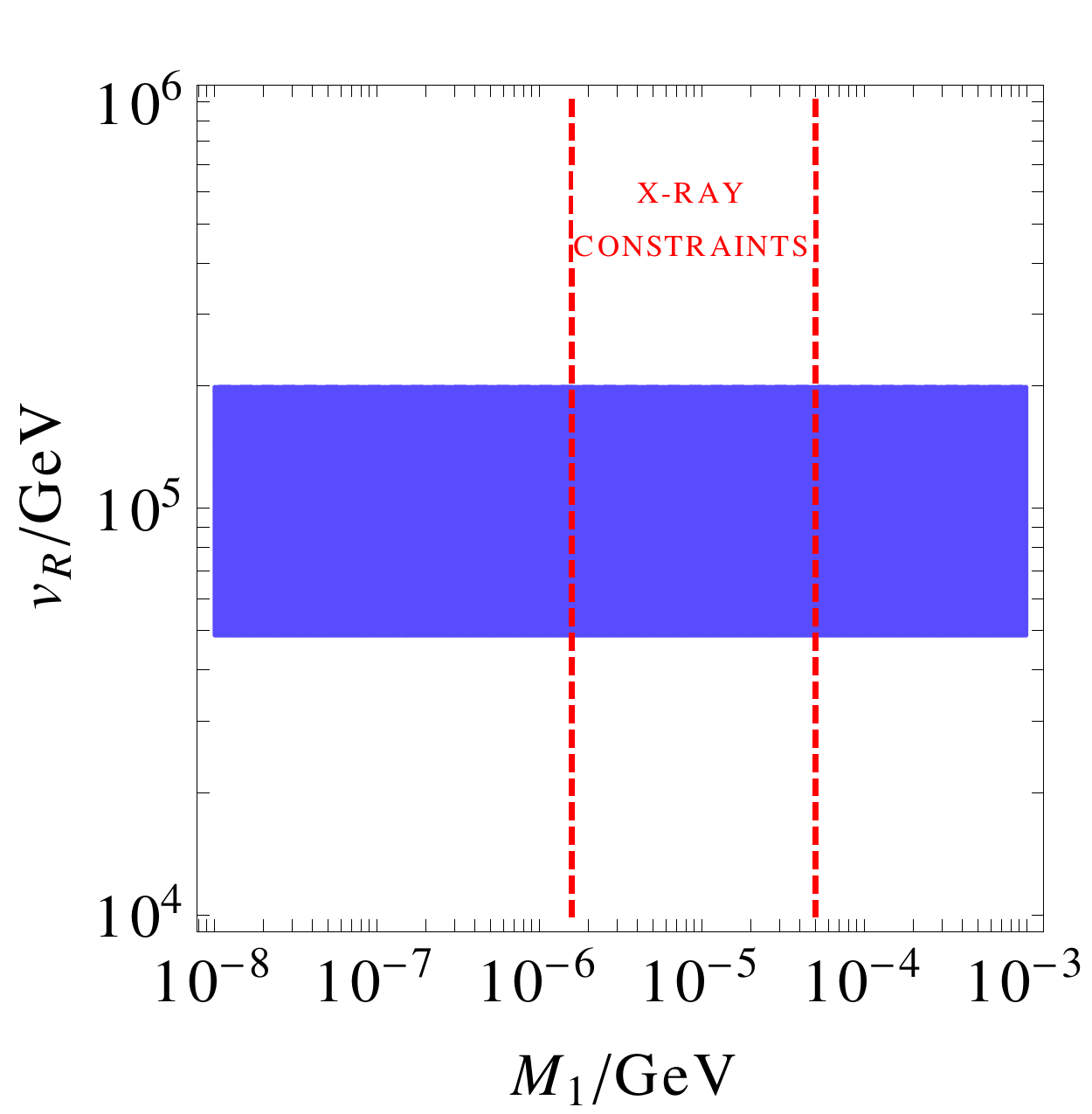} &
    \includegraphics[width=0.33\textwidth,angle=0]{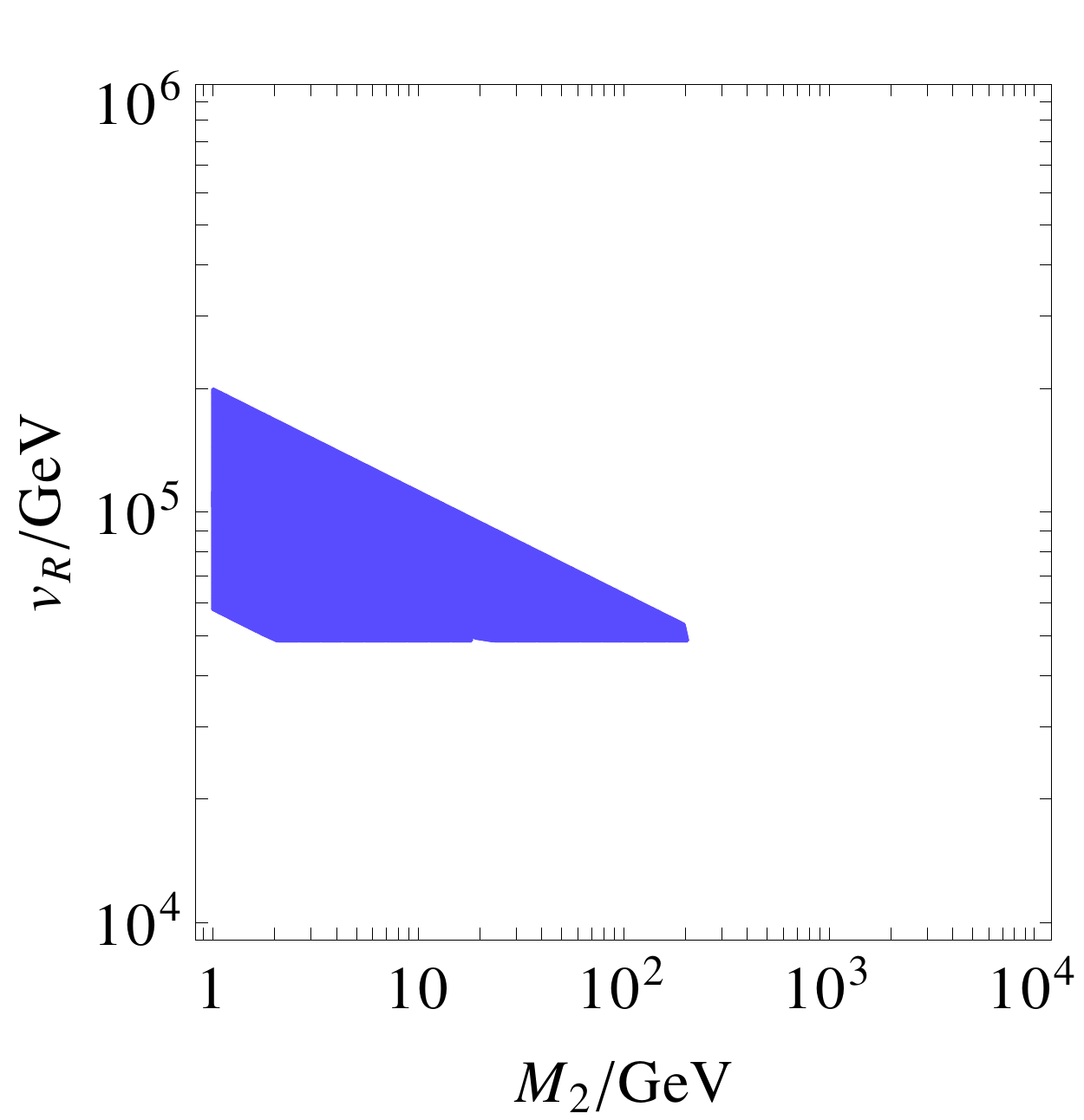}&
\includegraphics[width=0.33\textwidth,angle=0]{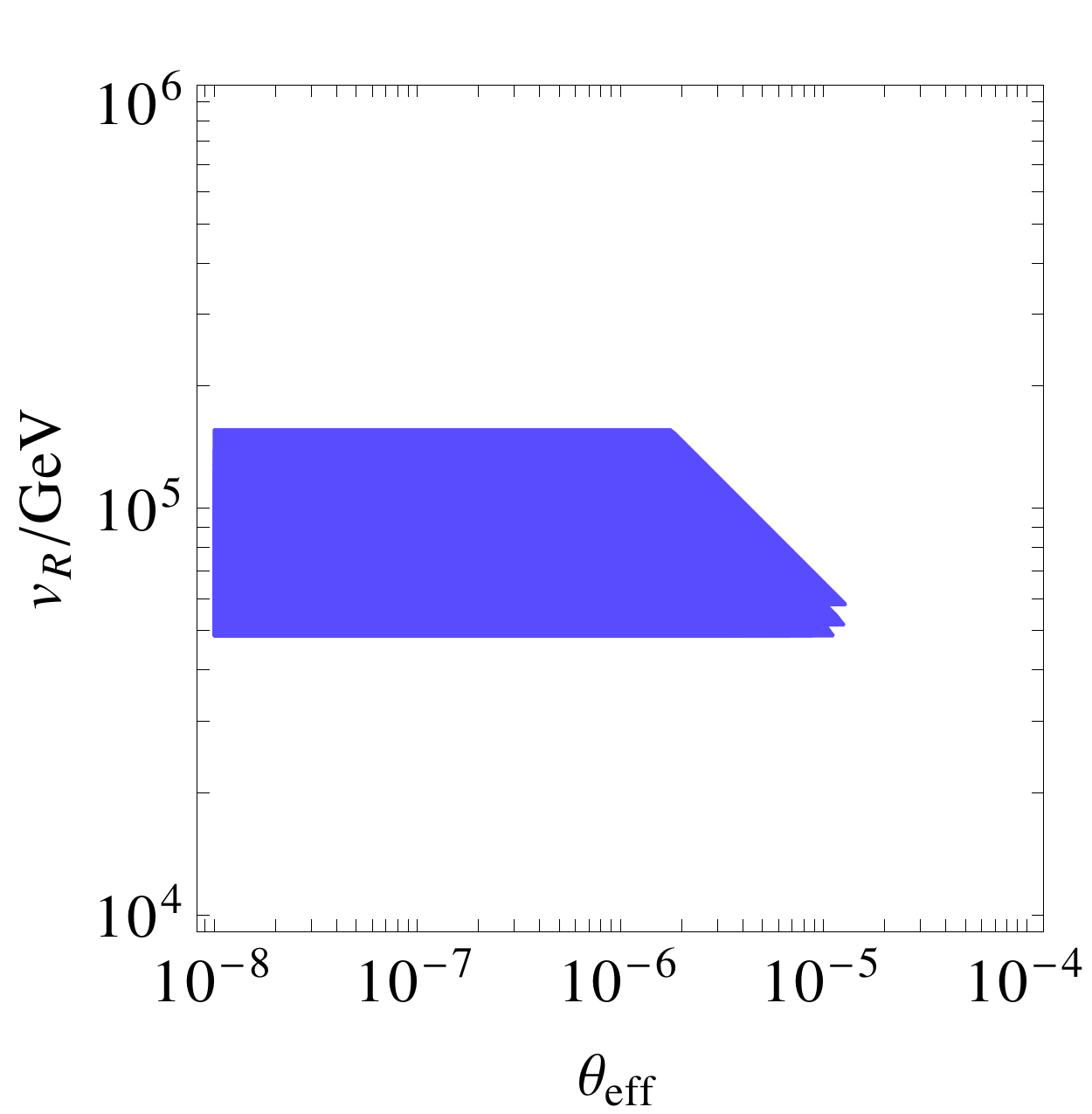}
 \end{tabular}
\begin{tabular}{ccc}
  \includegraphics[width=0.33\textwidth,angle=0]{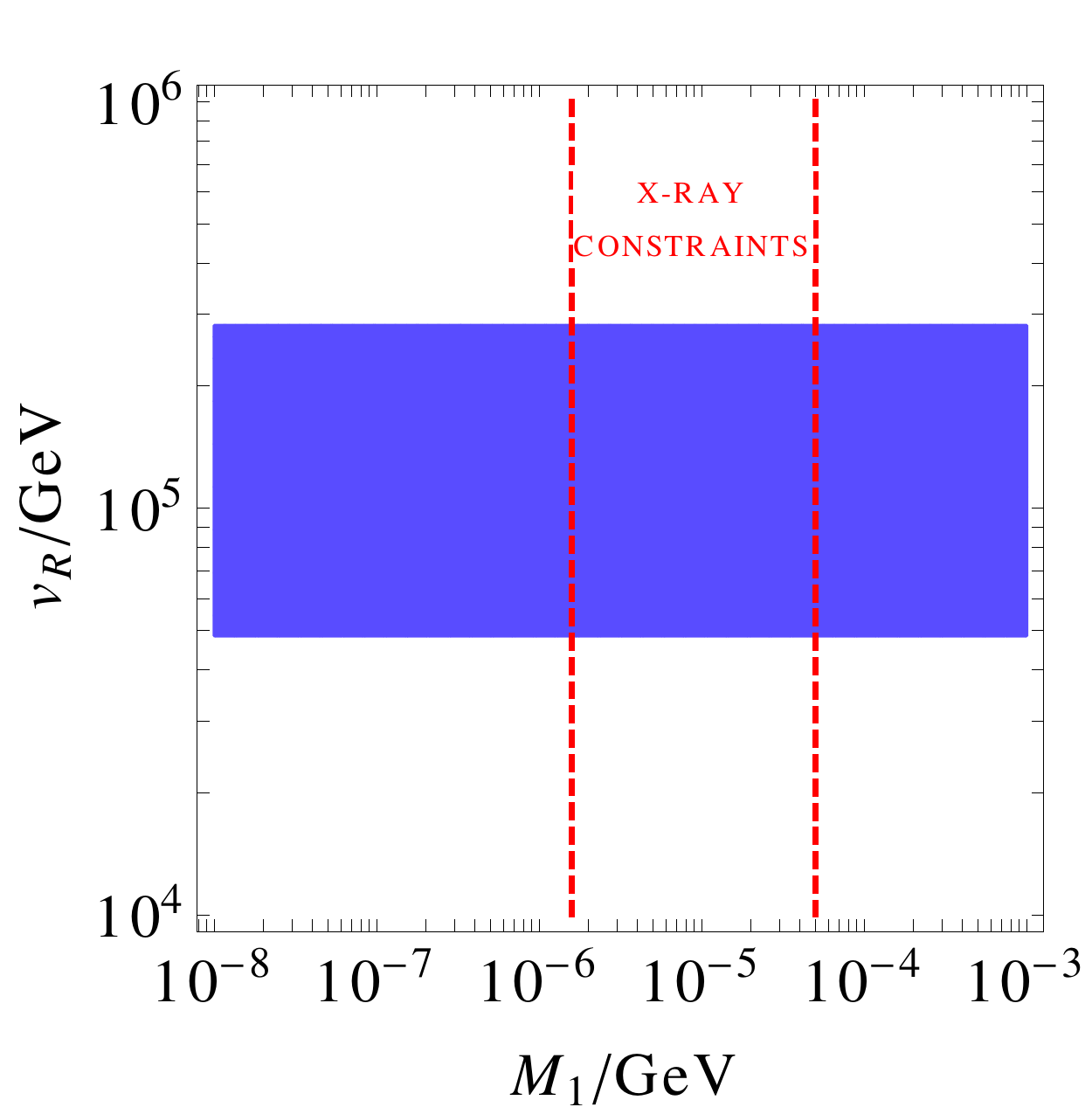} &
    \includegraphics[width=0.33\textwidth,angle=0]{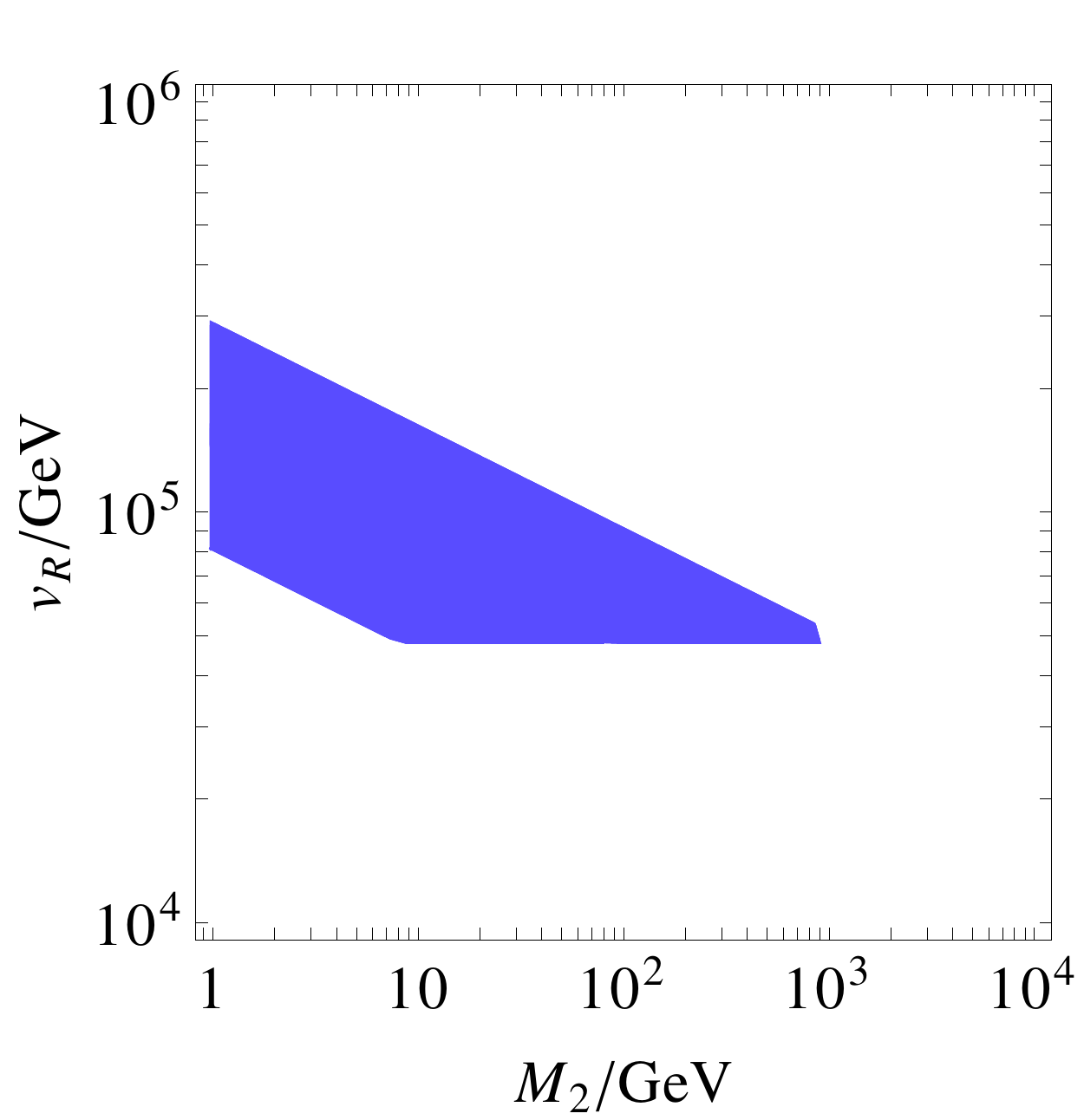}&
\includegraphics[width=0.33\textwidth,angle=0]{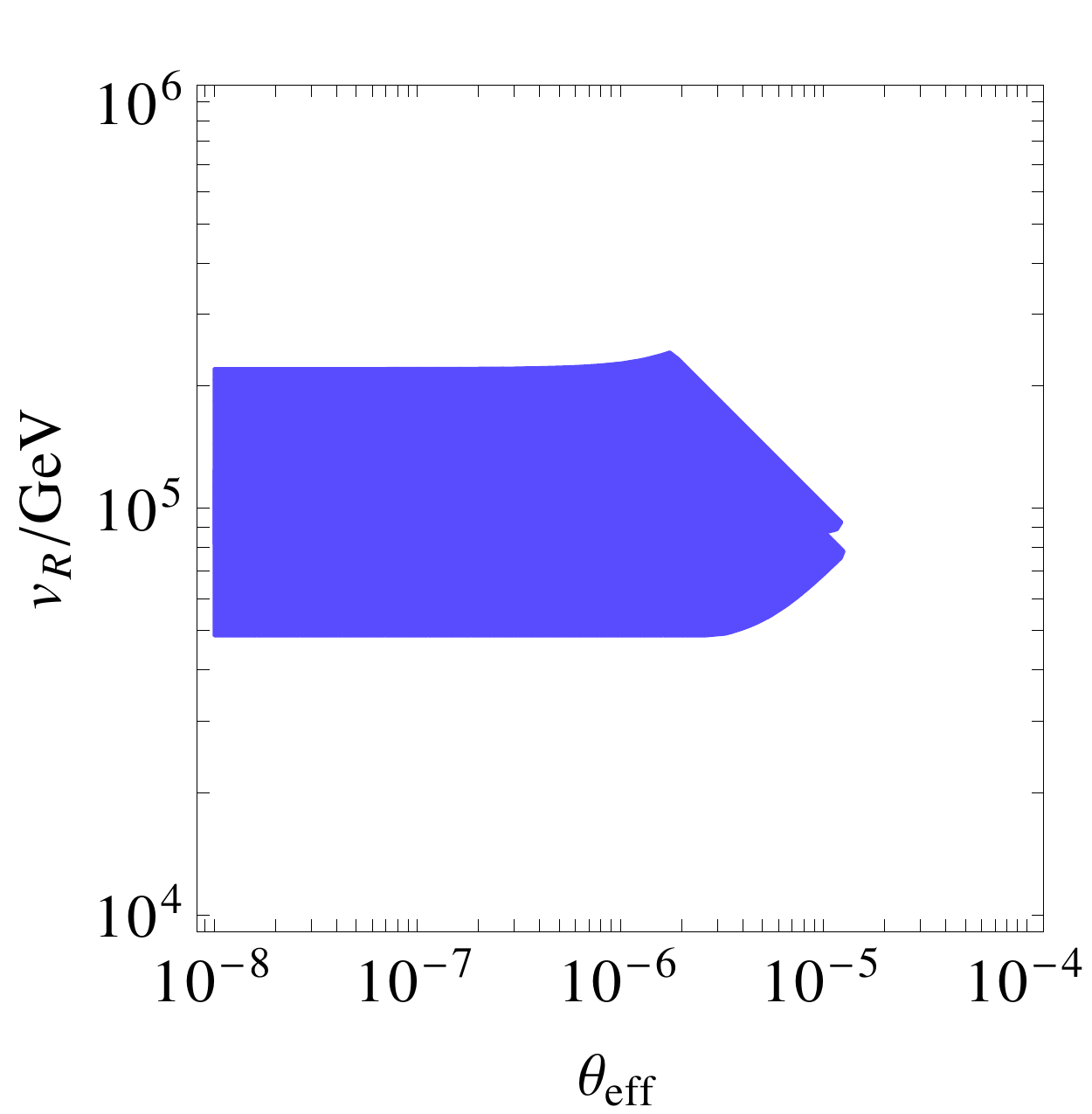}
 \end{tabular}
\caption{\label{figmixed}
\textbf{Mixed scenario.} The shaded region represents the future sensitivity of the next-to-next generation of $0\nu\beta\beta$ decay
experiments projected onto the $v_R-M_1$ plane (left panel), $v_R-M_2$ (central panel) and
$v_R-\theta_{eff}$ (right panel) when the decay rate is dominated by the $W_L-W_R$ and $W_R-W_R$ 
contributions with $\xi=0$ (upper panels) and $\xi=M^2_{W_L}/M_{W_R}^2$ (lower panels). In the analysis all the
relevant contributions have been simultaneously included. The bounds on $M_{W_R}$~\cite{Aad:2012ej},
the active-``heavy'' mixing~\cite{Atre:2009rg} and the X-ray constraints~\cite{Watson:2006qb} have been
also included. The PMNS angles and oscillation mass-squared differences have been fixed to the
central values given in Ref.~\cite{GonzalezGarcia:2012sz} and $m_1=10^{-2}$ eV, while the CP-phases of $U_{pmns}$ and $V$ have
been set to zero.}
\end{figure}

In order to be consistent with the rest of this work, associated with the absence of cancellation
in the light active neutrino contribution, we will focus in this section on the limit $\frac{v_L}{v_R}\rightarrow0$,
neglecting the first term of the $W_L-W_L$ contribution. As expected, we have checked
that if that term is switched on in the analysis, cancellations between the two terms of the $W_L-W_L$
contribution can perfectly occur for some part of the parameter space leading to better future
sensitivities to the $W_L-W_R$ and $W_R-W_R$ contributions. On the other hand,
we will not study the possibility of having a NP signal from the $W_L-W_L$ channel mediated by the
RH neutrinos since it would also require some level of fine tuning as demonstrated in Ref.~\cite{Blennow:2010th} for the type-I
seesaw case.

The left and central panel of Fig.~\ref{figmixed} show the sensitivity of the next-to-next generation 
of $0\nu\beta\beta$ decay experiments ($10^{-2}$ eV $<\lvert m_{\beta\beta}\rvert< 0.38$ eV) to the parameters of the model
by including all the relevant contributions and requiring the $W_L-W_R$ and $W_R-W_R$ contribution
to the $0\nu\beta\beta$ decay rate (second and third term in Eq.~(\ref{mbb3})) to be at
least 10 times larger than the $W_L-W_L$ contribution (first term of Eq.~(\ref{mbb3})) and 
$\xi=M^2_{W_L}/M_{W_R}^2$. The allowed region is projected onto the
$v_R$-$M_1$ plane (left panel) and $v_R$-$M_2$ plane (central panel). We have assumed that $U_{pmns}$ and
$V$ are real and fix the light neutrino mass scale to $m_1=10^{-2}$ eV. As in the previous plots, the
bounds on the $W_R$ mass~\cite{Aad:2012ej} and the active-``heavy'' mixing have been
included~\cite{Antusch:2006vwa,FernandezMartinez:2007ms,Atre:2009rg}. The X-ray
constraints~\cite{Watson:2006qb}, which apply if $N_1$ is the DM, are also
shown in Fig.~\ref{figlight1}: the region between the red dashed lines is ruled out. We have
used Eq.~(\ref{theta2}) in order to be consistent with the light neutrino mass and mixing pattern
as mentioned before. However, in order to illustrate better the impact of the mixing
$\theta$, we also show in Fig.~\ref{figmixed} (right panel) the sensitivity of the next-to-next
generation of $0\nu\beta\beta$ decay experiments to $v_R$ and an effective mixing $\theta_{eff}$
defined in the following way. We have assumed in Eq.~(\ref{mbb3}) that
$\sum_{i=2}^3 \left(\theta V\right)^*_{ei} V_{ei}\approx \theta_{eff}$ and $\sum_{i=2}^3
\left(\theta V\right)_{ei}^2 M_i \approx \theta_{eff}^2 M$, which is not true in general but a natural
assumption if no particular cancellations are involved. The results for $\xi=0$ are shown in the 
upper panels  while in the lower panels the $W_L-W_R$ mixing is maximal ($\xi=M^2_{W_L}/M_{W_R}^2$).  
The mixing plays a role only if it is close to the upper bound and
even in that case the future sensitivity is similar to the $\xi=0$ 
limit, as it can be observed in Fig 6.

Comparing Figs.~\ref{figlight1} and \ref{figmixed}, it is clear that the sensitivity to $v_R$
in this case and the light regime is very similar .
The main difference comes from the role of the $W_L-W_R$ channel, which in the light regime is completely
irrelevant but can be significant in the mixed scenario for values of the $W_L-W_R$ mixing close to the 
theoretical bound $\xi=M^2_{W_L}/M_{W_R}^2$. In fact, saturating the bound, the $W_L-W_R$ contribution
can be the dominant one in the region $\theta_{eff}\sim 10^{-7}$-$10^{-5}$ but for smaller values
of $\theta_{eff}$ it is negligible. This is easy to understand since the $W_L-W_R$ contribution
is proportional to $\theta$ while that of the $W_R-W_R$ channel do not depended on $\theta$. A
signal due to the $W_L-W_R$ channel is possible for such a small values of $\theta$ thanks to the enhancement
from $\langle p\rangle$ due to chirality argument and only if $\xi$ is close to its upper bound. On the other hand, the reason why Fig.~\ref{figmixed}
shows that the $v_R$ sensitivity is independent of $M_1$ is because for a large part of the parameter space
the $W_R-W_R$ channel dominates, where the $N_2$ and $N_3$ contribution is important.
In summary, a NP $0\nu\beta\beta$ decay signal can be expected for a quite small
light-sterile neutrino mixing ($\theta_{eff}\lesssim 10^{-5}$) if one of the RH
neutrinos is lighter than $1$ MeV while the rest are above the $0\nu\beta\beta$ decay scale. This
is interesting regarding the possibility of accommodating the DM in the left-right symmetric models as 
we will see in the next section.

\section{Complementary constraints}

In this section we will study the impact of the 1-loop corrections on the light neutrino masses and
whether the part of the parameter space which can be probed in 
future $0\nu\beta\beta$ decay experiments, as described above, is accessible by other experiments.

\subsection{1-loop corrections}

Since, in the scenarios studied here, the light neutrino contribution to the $0\nu\beta\beta$ decay rate
is suppressed with respect to the NP ones, one may expect that this significant NP lepton number violation 
contribution to the $0\nu\beta\beta$ decay rate could induce non-negligible 1-loop corrections to the 
light neutrino masses. Of course, if the 1-loop corrections are larger or similar to the tree-level 
contribution, they should be included in the analysis, which would modify our previous conclusions. The leading 
1-loop correction to the light neutrino masses is given by~\cite{Pilaftsis:1991ug,Grimus:2002nk}

\begin{equation}
\label{finitos}
\left(\delta M_{L}\right)_{\alpha\beta}= 
\frac{1}{(4 \pi v)^2}\left(\tilde{m}_D^T\right)_{\alpha i} \tilde{M}_i
\left\{\frac{3\ln\left(\tilde{M}_i^2/M^2_Z\right)}{\tilde{M}_i^2/M^2_Z-1}+\frac{\ln\left(\tilde{M}_i^2/M^2_H\right)}{\tilde{M}_i^2
/M^2_H-1}\right\}\left(\tilde{m}_D\right)_{i\beta}\,,
\end{equation}
where $\tilde{m}_D$ and $\tilde{M}=\text{diag}\left(M_1,M_2,M_3\right)$ are the Dirac and Majorana 
sub-matrices respectively, written in the basis in which the Majorana sub-matrix is diagonal, 
$M_Z$ is the mass of the $Z$ boson and $M_H$ the Higgs boson mass. Notice that the self energy diagrams
with $W_{L,R}$ bosons in the loop do not give any correction to the light neutrino 
masses since it is proportional to the external momentum~\cite{Grimus:2002nk}. The contribution would have been sensitive to 
$\xi$ and $M_{W_R}$, had $W_{L,R}$ contributed to the light neutrino corrections. Assuming that there is no fine tuning and the Yukawa couplings are of the 
same order, we can roughly estimate the size of the 1-loop corrections given by Eq.~(\ref{finitos}) 
as:
\bea
\delta M_{L}/m_{\nu} &\sim& 3\left(\frac{M_z}{4\pi v}\right)^2\ln\left(M_i^2/M_Z^2\right)+
\left(\frac{M_H}{4\pi v}\right)^2\ln\left(M_i^2/M_H^2\right),\; \text{for} \; M_{i}\gg M_Z,M_H, 
\nonumber \\
\delta M_{L}/m_{\nu} &\sim& \left(\frac{M_i}{4\pi v}\right)^2\left(3\ln\left(M_i^2/M_Z^2\right)+
\ln\left(M_i^2/M_H^2\right)\right), \; \text{for} \; M_{i}\ll M_Z,
\label{estimation}
\eea
Using the estimation given by the first equation above, we can conclude that, if $M_i\gg M_Z,M_H$,
the 1-loop corrections to the light neutrino masses are under control for the range of values that can 
be probed in future $0\nu\beta\beta$ decay experiments. In fact, for $M_i\sim 1$ TeV 
we have $\delta M_{L}/m_{\nu} \sim 10^{-2}$, and $\delta M_{L}/m_{\nu}$ gets smaller for smaller values of $M_i$;
for example, $\delta M_{L}/m_{\nu} \ll 10^{-4}$ for $M_i\ll M_Z$. This can be understood as follows. 
From Eq.~(\ref{finitos}), one can infer that the tree-level contribution 
is bigger than the loop induced ones because it has a similar structure to $m_D M_i^{-1} m_D^T$ 
but without the loop suppression, $1/(16 \pi^2)$. This is correct unless some cancellation is 
at work for the tree-level contribution, which is not the case studied here. Notice that, in this sense, 
the assumptions made in order to obtain Eq.~(\ref{estimation}) are quite reasonable. 

Therefore, we can conclude that the one-loop corrections to the light neutrino masses are negligible
and not relevant in our analysis. The lepton number violation source of the dominant NP contributions studied 
in the previous sections is the Majorana mass term generated dynamically for the RH neutrinos. 
Indeed, this source of lepton number violation is related to the light neutrino masses 
through the seesaw mechanism, and this correlation has been taken into account in the previous analysis. 
The dominant NP contribution to the $0\nu\beta\beta$ decay coming from the 
$W_R-W_R$ channel (or the $W_L-W_R$ channel in the mixed scenario) requires the suppression of the
standard (and long range) light neutrino one. Since we are not facing the possibility of having any cancellation 
in the light neutrino contribution, in order to achieve this suppression the Yukawa couplings 
and $v_L/v_R$ should be small. The $W_R-W_R$ contribution can be dominant because the RH mixing is not 
constrained in contrast with the active-heavy mixing $\theta$, which is necessarily small as the Yukawa
couplings. 
The $W_L-W_R$ channel can dominate in the mixed scenario (only for large
$\xi$) due to the enhancement coming from the NME and the linear dependence on the active-heavy 
mixing $\theta$.

\subsection{Other experimental bounds}

The charged LFV experiments are also sensitive 
to the parameters of the model that can be probed in $0\nu\beta\beta$ decay experiments. Among them,
$\mu\rightarrow e \gamma$, $\mu\rightarrow 3 e$ and $\mu \rightarrow e$ conversion give the stronger
bounds. First of all, the small active-heavy mixing required here in order to have a significant 
NP contribution to $0\nu\beta\beta$ decay ($\theta\lesssim 10^{-5}$), renders the type-I seesaw like 
contribution to this processes completely negligible since the strongest present bound coming from 
$\mu\rightarrow e \gamma$ gives the constraint $|(\theta^\dagger\theta)_{e\mu}|\lesssim 10^{-5}$. 
A complete calculation of the charged LFV branching ratios in the MLRSM can be found in~\cite{Tello_thesis:2012}. The most relevant constraint in the context of this
work comes from $\mu\rightarrow e \gamma$, whose branching ratio, to zeroth order on $\theta$ and $\xi$,
is given by:

\be
Br_{\mu\rightarrow e \gamma}\approx 2.6 \times 10^{-10}\left(\frac{\text{TeV}}{M_{W_R}}\right)^4
\left( \frac{|\left(M_RM_R^*\right)_{\mu e}|}{M_{W_R}^2}\right)^2,\label{muegamma}
\,
\ee
for $M_{\Delta_{L,R}}\gg M_{W_R}$. Applying the present experimental constraint~\cite{Adam:2013mnn} 
to Eq.~(\ref{muegamma}), the bound on $M_R$ reads

\be
M_R \lesssim \left(\frac{M_{W_R}^2}{4.6 \text{TeV}}\right).
\label{boundMR}
\ee
Saturating the lower bound on $M_{W_R}$, one obtains $M_R\lesssim 1$ TeV. This is the 
largest $M_R$ that can be probed with $0\nu\beta\beta$ decay experiments as it can be
seen in Fig.~\ref{figheavy} (it corresponds to the bottom right corner of the shaded regions
in the left panels). This means that the future $\mu\rightarrow e \gamma$ experiments can
be sensitive at least to that corner of the parameter space, which can 
also give a signal in $0\nu\beta\beta$ decay experiments. One should, however, keep in mind that
the flavour structure of $M_R$ plays an important role, being $\mu\rightarrow e \gamma$
experiments indeed sensitive to $\left(M_RM_R^*\right)_{\mu e}$ and $0\nu\beta\beta$ decay 
mainly to $\left(M_R\right)_{ee}$ if the heavy neutrino spectrum is hierarchical. 
Therefore, a NP signal in $0\nu\beta\beta$ decay experiments does not necessarily imply also a 
signal in future $\mu\rightarrow e \gamma$ experiments. On the other hand, the bound in 
Eq.~(\ref{boundMR}) has been extracted assuming that $M_{\Delta_{L,R}}\gg M_{W_R}$, but 
smaller masses of the triplets can clearly enhance the branching ratio~\cite{Tello_thesis:2012}. 
The same applies for the $\mu\rightarrow 3 e$ and $\mu \rightarrow e$ conversion case since their 
branching ratios are inversely proportional to the triple masses. Therefore, we can not extract 
a bound like Eq.~(\ref{boundMR}) from  $\mu\rightarrow 3 e$ and $\mu \rightarrow e$ conversion, 
since for large triplet masses the branching ratios are very suppressed. 

So far, in our LFV analysis we have neglected the $\xi$ contribution. If one switches on the 
left-right mixing $\xi$, the following constraint from $\mu\rightarrow e \gamma$ can be
extracted~\cite{Barry:2013xxa}:

\be
|\left(m_D\right)_{\mu e}|\xi \lesssim 2\,\text{KeV},
\ee
which can be roughly translated into $\theta \xi \lesssim \,\frac{2\,\text{KeV}}{M_R}$, which
is basically compatible with most of the parameter space that can give a NP signal in 
$0\nu\beta\beta$ decay, since $\xi<10^{-3}$ and $\theta<10^{-5}$. Only if $M_R$ is close to
the TeV and $\xi$ saturates the present bound ($\xi\leq M_{W_L}/M_{W_R}<10^{-3}$), could the 
mixing $\xi$ have an impact in $\mu\rightarrow e \gamma$. Basically, we could probe the 
same part of the parameter space commented above but for values of $\xi$ close to its present
bound, again with the important warning that the flavour structure plays an essential role here. 

Finally, in the MLRSM the electric dipole moment (EDM) can be considerably 
enhanced with respect to the SM result (up to 10 orders of magnitude). This is because the 
SM contribution to the EDM appears at four loops while the left-right symmetric model can provide a huge enhancement 
due to the left-right mixing $\xi$~\cite{Nieves:1986uk}. In fact, the EDM experiments
can be sensitive in the future to part of the parameter space studied here~\cite{Nemevsek:2012iq}, 
mainly through the imaginary part of $\left[\left(m_D\right)_{ee}\xi\right]$.

\section{Dark Matter}

In this section we study the possibility of having a successful DM candidate in the context of the
MLRSM when the $0\nu\beta\beta$ decay rate is dominated by NP contributions.
The first question which arises from the results of the previous section is whether $N_1$ can be DM in the light regime,
namely with mass $M_1 \lesssim \mathcal{O}(\rm{MeV})$.
This reminds us of the Dodelson-Widrow (DW) scenario~\cite{Dodelson:1993je}, where a KeV neutrino is
produced via neutrino oscillations and can be a viable DM candidate\footnote{A recent study for KeV-neutrino DM on the $0\nu\beta\beta$ decay
in the context of the type-I seesaw can be found in Ref.~\cite{Merle:2013ibc}, where the KeV neutrino contribution to the $0\nu\beta\beta$ rate is subleading due to the X-ray bound.}. In the left-right symmetric
models, however, a RH KeV neutrino $N_1$ would be thermally produced via the $W_R$ or $Z_R$ exchange and
decouples from the thermal bath at the freeze-out temperature $T_f$,

 \begin{equation}
 T_f \sim 400 \,\,\mbox{MeV}
 \left( \frac{g_{*}(T_f)}{70} \right)^{1/6} \left( \frac{M_{W_R}}{ 5\,\rm{TeV}} \right)^{4/3},
 \end{equation}
 where $g_{*}(T_f)$ is the number of relativistic degrees of freedom at freeze-out. The rule of thumb to
estimate $T_f$ is to set the interaction rate equal to the expansion rate of the Universe. Given that we
are interested on the region of the parameter space in which the NP dominates the $0\nu\beta\beta$ decay
rate, the $W_R$ mass should be in the range $M_{W_R}\sim 1$-$15$ TeV (see Figs.~\ref{figmixed}). Therefore,
$N_1$ is highly relativistic ($M_1\lesssim$ MeV $\ll T_f$) at freeze-out and the resulting relic density
is~\cite{Nemevsek:2012cd},

\begin{equation}
\Omega_{N_1} \simeq 3.3\, \left( \frac{M_1}{1 \mbox{KeV}} \right) \left( \frac{70}{g_{*}(T_f)} \right),
 \end{equation}
which, for $M_1 \sim \rm{KeV}$, would be much larger than the observed DM relic density
$\Omega_{\rm{DM}}=0.265$~\cite{Ade:2013zuv}. This constraint is much severer than the X-ray constraints shown in Fig.~\ref{figlight1}
and the Big Bang Nucleosynthesis (BBN) bound, which is indeed still compatible at $\sim2\sigma$ with the
existence of one extra relativistic species~\cite{Mangano:2011ar,Hamann:2011ge}.
A possible way out has recently been proposed and studied in detail in Refs.~\cite{Bezrukov:2009th,Nemevsek:2012cd}. Basically, the idea is to
dilute the number density of $N_1$ by the injection of entropy into the thermal bath after $N_1$ freezes out.
To be more specific, the out-of-equilibrium decay of $N_2$ and/or $N_3$, of mass around GeV, into SM particles can increase
the entropy of the Universe, leading to faster Universe expansion and in turn a smaller $N_1$ density.
The set of constraints that should be satisfied if $N_1$ as DM was once in thermal equilibrium has been
summarized in Ref.~\cite{Bezrukov:2009th}. In particular, the authors claim that the required
entropy injection can be achieved if $M_{W_R}\gtrsim 10$-$16$ TeV while the Lyman-$\alpha$ constraints
require $M_1\gtrsim1$ KeV. It turns out that these bounds and the rest of the constraints listed in
Ref.~\cite{Bezrukov:2009th} are compatible with a future NP signal in $0\nu\beta\beta$ decay experiments described in
Sec.~\ref{mixedregime}. Notice that the RH neutrino spectrum required to have DM
($M_1\sim$ KeV and $M_2,M_3\sim 1$-$10$ GeV) belongs to the mixed scenario where
a NP signal in $0\nu\beta\beta$ decay experiments is possible. It should be remarked that
the constraint $M_{W_R}\gtrsim 10$-$16$ TeV is in obvious tension with a future NP
signal in the $0\nu\beta\beta$ decay. However, as we have mentioned in the previous section, if $v_L$
is switched on in the above analysis a cancellation between the two terms in the light neutrino contribution
can take place such that a NP signal can be possible for $M_{W_R}\gtrsim 16$ TeV.

On the other hand, in Ref.~\cite{Nemevsek:2012cd} an alternative scenario able to relax the bound
on $M_{W_R}$ from Ref.~\cite{Bezrukov:2009th} was carefully analyzed. In this scenario the
desired dilution of the number density of $N_1$ is achieved for
$M_1 \simeq 0.5$ KeV, $M_{2} \sim 140$ MeV and $M_{3} \sim 245$ MeV,
with $M_{W_R}\sim 5$ TeV and the help of a particular right-handed flavor structure such
that $N_2$'s coupling constant to SM leptons is stronger than that of $N_1$. We refer the readers
to Ref.~\cite{Nemevsek:2012cd} for the details of the analysis. In any case, as it was already
pointed out in Ref.~\cite{Nemevsek:2012cd}, the contribution to the $0\nu\beta\beta$ decay rate
from the $W_R-W_R$ channel associated with such spectrum can be testable in the future $0\nu\beta\beta$
decay experiments as we have confirmed in Sec.~\ref{mixedregime}. However, we would like to remark that the
$W_L-W_R$ contribution can also be very relevant in this case, as it was explained in the previous section.

Finally, in the left-right symmetric models, in principle the neutral component of the right-handed triplet
$\Delta_R^0$, which is a singlet under the SM gauge group, could also be a DM candidate.
However, it decays at one-loop into two photons via $W_R$ exchange~\cite{Nemevsek:2012cd},
i.e.,

 \begin{equation}
 \Gamma_{\Delta_R^0 \rightarrow \gamma\gamma} \sim 10^{-52}\,\, \mbox{GeV} \left( \frac{m_\Delta}{\mbox{KeV}} \right)^3
 \left( \frac{10^{13} \, \mbox{GeV}}{ M_{W_R} } \right)^2.
 \end{equation}
The X-ray constraints on KeV DM resulting from observations on galaxies and clusters of
galaxies~\cite{Abazajian:2001vt} requires $\tau_{\Delta_R^0 \rightarrow \gamma\gamma}=1/
\Gamma_{\Delta_R^0 \rightarrow \gamma\gamma} \gtrsim 10^{28}$ sec or $\Gamma_{\Delta_R^0
\rightarrow \gamma\gamma} \lesssim 10^{-52}$ GeV. Therefore, a KeV $\Delta_R^0$ would
imply a too heavy $W_R$ such that the contribution of the NP channels involving $W_R$
to the $0\nu\beta\beta$ decay would be completely negligible. Nevertheless, the X-ray constraints apply only
to DM with masses around $1$-$20$ KeV. In principle this leaves
another window of $\Delta_R^0$ mass which can be studied. However, other constraints
make this possibility quite unfeasible. First, the mass of DM is
constrained to be larger than KeV~\cite{Viel:2007mv} because of the Lyman-$\alpha$ observations.
Second, for $M_{\Delta_R^0} \gtrsim 20$ KeV,
$\tau_{\Delta_R^0}$ still has to be longer than the age of the Universe, around $10^{18}$ sec,
which results again in a very heavy $W_R$ that renders any NP contribution to $0\nu\beta\beta$ decay
far beyond the future experimental sensitivity.

In summary, in spite of the existence of various constraints, the left-right symmetric models can accommodate a
KeV RH neutrino as a successful DM candidate which can lead to a NP signal in the next-to-next
generation of $0\nu\beta\beta$ decay experiments.

\section{Conclusions}

We have studied the $0\nu\beta\beta$ decay phenomenology in the MLRSM. In particular,
we have analyzed under which conditions a $0\nu\beta\beta$ decay signal can come mainly from NP contributions
associated with this model. Special attention has been paid to the correlation among the different NP
contributions and the standard light neutrino one. This correlation emerges from the neutrino mass
generation mechanism and should always be considered in the analysis. The scenario in which an accidental cancellation in the $W_L-W_L$ contribution takes place
has not been explored. The role of the $W_L-W_R$ mixing $\xi$ and 
the possibility of having a NP dominated $0\nu\beta\beta$ decay signal compatible with DM is also investigated.

We have distinguished three different regions of the parameter space based on the mass of the RH neutrinos
: (i) all the masses heavier than GeV, denoted by heavy regime; (ii) masses lighter than MeV,
dubbed light regime; (iii) the lightest mass below the MeV and the rest above GeV, called mixed scenario.
Notice that (i) has been extensively studied in the literature, but (ii) and (iii) have not been 
analyzed before in detail for the left-right symmetric models (at least the $0\nu\beta\beta$ decay phenomenology).

In the heavy region, we have found that the dominant NP contribution emerges mainly from the $W_R-W_R$
channel mediated by the heavy neutrinos. To be more precise,
it has been shown that this dominant NP contribution could be measured in the next-to-next generation of
$0\nu\beta\beta$ decay experiments for the window $M_{W_R}\sim 1-15$ TeV ($M_{W_R}\sim 1-20$ TeV), if the 
active-heavy mixing is smaller than $\sim 10^{-5}$ and the right-handed triplet ``Yukawa''
coupling satisfies $3\cdot10^{-6} \lesssim\left(Y_\Delta\right)_{ee}\lesssim 3\cdot 10^{-2}$ 
($3\cdot10^{-6} \lesssim\left(Y_\Delta\right)_{ee}\lesssim 8\cdot 10^{-2}$), which corresponds to a
range of heavy neutrino masses from GeV to TeV for $\xi=0$ ($\xi=M^2_{W_L}/M_{W_R}^2$). We have also shown that neglecting the present correlation
between the various contributions, and that with the light neutrinos in particular, can lead to incorrect
results. For instance, it is found that the region of the parameter space which can be experimentally
probed when only the $W_L-W_R$ contribution is included in the analysis shrinks considerably if all the
contributions are included at once and their correlations are not neglected. 

The results for the light region turn out to be similar to those of the heavy region. We have found that a future $0\nu\beta\beta$ decay
NP signal can come only from the $W_R-W_R$ channel since the $W_L-W_R$ contribution cancels out in this regime.
In particular, we have shown in which part of the parameter space this is possible and we found a
similar but weaker sensitivity of the next-to-next generation of $0\nu\beta\beta$ decay experiments:
$M_{W_R}\lesssim 8$ TeV ($M_{W_R}\lesssim 12$ TeV) for $\xi=0$ ($\xi=M_{W_L}^2/M_{W_R}^2$). A NP signal can 
be expected for $\left(M_R\right)_{ee}\sim15$ KeV$-$ MeV ($\left(M_R\right)_{ee}\sim$ KeV$-$ MeV)
for $\xi=0$ ($\xi=M_{W_L}^2/M_{W_R}^2$), while the triplet Yukawa coupling should be inside the 
region $10^{-10} \lesssim\left(Y_\Delta\right)_{ee}\lesssim 10^{-8}$. This uncomfortably small value of $Y_\Delta$ is required
in order to have very light RH neutrinos ($M_i<1$ MeV) since $M_i\sim Y_\Delta v_R$,
that seems unnatural but might be achieved with the help of an additional global symmetry. On the
other hand, notice that in this regime the $W_L-W_L$ contribution is proportional to $v_L/v_R$ and
therefore a small value of $v_L$ ($v_L \lesssim 0.07$ GeV) guarantees a dominant $W_R-W_R$ contribution
to the $0\nu\beta\beta$ decay rate. If $v_L/v_R\gg (M_{W_L}/M_{W_R})^4$, which is still experimentally
allowed, the RH neutrinos can dominate the process via the $W_L-W_L$ channel,
contrary to the type-I seesaw case in which the decay rate is very suppressed if all the
RH neutrinos are lighter than the $0\nu\beta\beta$ decay scale.

In the intermediate regime, with RH neutrinos in both regions ($M_1\lesssim1$ MeV and
$M_2,M_3\gtrsim1$ GeV), if the $W_L-W_R$ mixing is close to the theoretical upper bound $\xi=M^2_{W_L}/M_{W_R}^2$, 
the role of the $W_L-W_R$ channel can be relevant, in contrast with the previous cases. We have 
found that a NP signal coming from the $W_L-W_R$ channel could take place for $M_{W_R}\sim 1-10$ TeV and
an active-heavy neutrino mixing $\theta\sim 10^{-7}$-$10^{-5}$.  Indeed, a signal from the $W_R-W_R$ 
channel can be expected in a larger region, even for smaller values of $\theta$ since its 
contribution is independent of the active-heavy neutrino mixing. In this case we have focused our study on the limit
$v_L/v_R\rightarrow 0$, but we have checked that if $v_L$ is switched on in the analysis the future
sensitivity to $v_R$ is much better. However, this can only take place when there 
is a cancellation between the type-I and type-II seesaw terms in the light neutrino contribution.

In order to study the impact of the $W_L-W_R$ mixing $\xi$, we have analyzed the following two extreme limits: 
$\xi=M_{W_L}^2/M_{W_R}^2$ (maximal) and $\xi=0$ (negligible). We have shown 
that the inclusion of the $W_L-W_R$ mixing $\xi$ can have some impact on the results but it is not very 
significant, with the possible exception of the mixed scenario where a large mixing is required in order 
to have a relevant role of the $W_L-W_R$ channel. In general, due to the enhancement on the $W_R-W_R$ contribution
for maximal mixing, the sensitivity to $M_{W_R}$ is about a factor $1.5$ larger for $\xi=M_{W_L}^2/M_{W_R}^2$ 
than for $\xi=0$ in all the regions under study.

We have also analyzed the role of the complementary bounds coming from charged LFV processes 
and the the impact of the 1-loop corrections to the light neutrino masses in the context of this work. It 
turns out that the light neutrino masses are stable under 1-loop corrections since they might be important
only if a cancellation takes places in the light neutrino masses, but not in the case studied here
where there is a general suppression of the light masses with small $v_L/v_R$ and the Yukawa couplings. Future charged LFV experiments
 might allow us to probe part of the parameter space that can be responsible for a NP signal in $0\nu\beta\beta$ decay experiments, but
only a small region in the heavy regime around $M_R\sim 1$ TeV (the bottom right corner of the shaded regions in the left 
panels of Fig.~\ref{figheavy}). In fact, a more complete study, beyond the scope of this work, including the 
effect of triplet masses close to their lower bounds, which can enhance the branching ratios, would be required in order to clarify the issue. A large left-right mixing $\xi$ can also be probed in future EDM experiments as it was shown in 
Ref.~\cite{Nemevsek:2012iq}.

Finally, the following DM-related question has also been addressed. Can a NP dominated $0\nu\beta\beta$
decay signal be compatible with a successful DM candidate in the left-right symmetric models?
We conclude that, regardless of the various strong constraints, it is still possible for the
scenario proposed in Ref.~\cite{Bezrukov:2009th}, where a KeV RH neutrino can be the
DM if the scale of the other heavy neutrinos is around $1-10$ GeV and $M_{W_R}\gtrsim 10$-$15$ TeV.
We have shown that the $0\nu\beta\beta$ decay signal can be induced by the RH neutrinos
through the $W_L-W_R$ and $W_R-W_R$ channel. Additionally, Ref.~\cite{Nemevsek:2012cd} opens a window of $M_{W_R}\sim5$ GeV within the horizon
of LHC after the QCD phase transition is carefully included.

\begin{acknowledgments}
We thank to S. Petcov and Javier Menendez for useful discussions and important remarks. This work was partially
supported by the ITN INVISIBLES (Marie Curie Actions, PITN- GA-2011-289442).  W.-C.H. would like to thank the hospitality of IFPA at Universit\'{e} de Li\`{e}ge, where part of this work was performed.
\end{acknowledgments}

\end{document}